\documentclass[singlecolumn,floatfix]{revtex4-2}
\usepackage{amsfonts}
\usepackage{amsmath}
\usepackage{amssymb}
\usepackage{amsthm}
\usepackage{array}
\usepackage{epsfig}
\usepackage{rotating,graphicx}
\usepackage{stix}
\usepackage{bm}%
\usepackage{verbatim}
\usepackage{tensor}
\usepackage{xcolor}
\usepackage{picinpar}
\usepackage{bm}
\usepackage{longtable}
\usepackage{multirow}
\theoremstyle{plain}
\usepackage{amsthm}

\newenvironment{aside}{
\begin{quotation}
\begin{footnotesize}
}
{
\end{footnotesize}
\end{quotation}
}

\theoremstyle{definition}

\newtheorem{exatitle}{Example}

\newenvironment{myexample}[2]%
{\begin{exatitle} \label{#2} #1 \end{exatitle}}%
{\hfill $\Box$ \\}

\newcommand{\ket}[1]{| #1 \rangle}
\newcommand{\bra}[1]{\langle #1 |}
\newcommand{\braket}[2]{\langle #1 | #2 \rangle}

\newcommand{\pket}[1]{[ #1 ]}

\newcommand{\td}{\text{d}}
\newcommand{\Tr}{\text{Tr}}

\newcommand{\eg}{\hbox{\em e.g.{}}}

\newcommand{\ie}{\hbox{\em i.e.{}}}
\newcommand{\wrt}{\hbox{w.r.t.{}}}

\newcommand{\rhs}{\hbox{r.h.s.{}}}

\newcommand{\Ps}{\mathbb{P}}

\newcommand{\evc}[1]{\lcurvyangle #1 \rcurvyangle}

\makeatletter
\g@addto@macro\bfseries{\boldmath}
\makeatother

\usepackage{bm}

\begin{document}

\title{Curves in quantum state space, geometric phases, and the brachistophase}

\author{C.{} Chryssomalakos}
\email{chryss@nucleares.unam.mx}
\affiliation{Instituto de Ciencias Nucleares \\
	Universidad Nacional Aut\'onoma de M\'exico\\
	PO Box 70-543, 04510, CDMX, M\'exico.}
	
\author{A.{} G.{} Flores-Delgado}
\email{ana.flores@correo.nucleares.unam.mx}
\affiliation{Instituto de Ciencias Nucleares \\
	Universidad Nacional Aut\'onoma de M\'exico\\
	PO Box 70-543, 04510, CDMX, M\'exico.}	
	
\author{E.{} Guzm\'an-Gonz\'alez}
\email{edgar.guzman@correo.nucleares.unam.mx}
\affiliation{Departamento de F\'{\i}sica \\
	Universidad Aut\'onoma Metropolitana-Iztapalapa\\
PO Box 55-534, 09340, CDMX, M\'exico.}

\author{L.{} Hanotel}
\email{khanotel@hse.ru}	
\affiliation{Tikhonov Moscow Institute of Electronics and Mathematics\\
HSE University\\
Tallinskaya ul.{} 34, 123592, Moscow, Russia.}

\author{E.{} Serrano-Ens\'astiga}
\email{ed.ensastiga@uliege.be}
\affiliation{Centro de Nanociencias y Nanotecnolog\'{\i}a \\
Universidad Nacional Aut\'onoma de M\'exico\\
PO Box 14, 22800, Ensenada, Baja California, M\'exico\\
 and
\\
Institut de Physique Nucléaire, Atomique et de Spectroscopie, CESAM
\\ Universit\'e de Liège
\\ B-4000 Liège, Belgium
}

\begin{abstract}
	\noindent
Given a curve in quantum spin state space, we inquire what is the relation between its geometry and the geometric phase accumulated along it. Motivated by Mukunda and Simon's result that geodesics (in the standard Fubini-Study metric) do not accumulate geometric phase, we find a general expression for the derivatives (of various orders) of the geometric phase in terms of the covariant derivatives of the curve. As an application of our results, we put forward the \emph{brachistophase} problem: given a quantum state, find the (appropriately normalized) hamiltonian that maximizes the accumulated geometric phase after time $\tau$ --- we find an analytical solution for all spin values, valid for small $\tau$. For example, the optimal evolution of a spin coherent state consists of a single Majorana star separating from the rest and tracing out a circle on the Majorana sphere.
\end{abstract}
\date{\today}
\maketitle
 
\tableofcontents
\newsavebox{\Potb}

\section{Introduction}
\label{Intro}
The geometric phase accumulated during the evolution of a quantum system plays an essential role in a variety of physical phenomena, such as the nuclear dynamics in the Born-Oppenheimer molecular theory~\cite{Mea.Tru:79,Mea:92}, physical properties of materials like polarization, magnetization, or in the various  Hall effects ~\cite{Res:94,Xia.Min.Niu:10,Sha.Wil:89,chr.jam:04,Co.La.Bo:19}, to mention but a few. Additionally, it has been proposed as a key ingredient in  the implementation of quantum computing through holonomic quantum gates~\cite{Zan.Ras:99}. Moreover, several sets of universal quantum gates include a unitary operation that imparts a generic geometric phase to a state~\cite{Nie.Chu:11}. The first formal deduction of geometric phase in the quantum realm was given by Michael Berry in 1984~\cite{Ber:84}, considering  a system in a non-degenerate hamiltonian eigenstate, in the adiabatic approximation.  Eventually, the concept was generalized to non-abelian (Wilczek-Zee) geometric phases~\cite{Wil.Zee:84}, nonadiabatic evolutions~\cite{Aha.Ana:87,Ana:88} and even in the case of non-cyclic curves~\cite{Sam.Bha:88}, reaching its most general form in the work of Simon and Mukunda in \cite{Muk.Sim:93}. Nowadays, there is experimental evidence of the geometric phase for both cyclic and non-cyclic curves~\cite{Bit.Dub:87,Zho.Mar.Mou:Mei:20}. The mathematical characterization of the geometric phase as a holonomy of a connection dictated by the Schr\"odinger evolution of the quantum state~\cite{Sim:83} underlies the generalizations mentioned above. Following this geometrical point of view, we ask the following question: given a curve in quantum state space, what geometrical properties of the curve give rise to the geometric phase? A key result,  within this mathematical framework, is that geodesic curves (in the natural Fubini-Study metric), \ie, curves without acceleration, do not accumulate geometric phase \cite{Muk.Sim:93}. Consequently, the geometric phase associated with a curve, parametrized by arclength, depends on the (covariant) derivatives of second and higher orders of the curve. Spelling out in detail this relation is one of the main goals of the present work.

The experimental generation of geometric phases in quantum computation faces multiple challenges, like  decoherence and other systematic errors~\cite{Fue.Gir.Liv:05},  necessitating  the implementation of quantum gates in the shortest possible time, within the most efficient scheme. Several scenarios have been studied in this context \cite{Fue.Gir.Liv:05,Sjo.Ton:12,Xu.Ton.Sjo:18}, exemplified in the well-known quantum brachistochrone problem~\cite{Wan.All.Jac.Llo:15}: realize a quantum gate, or ``control protocol'', in the shortest possible time under suitable conditions. We contribute to this application-oriented direction by first determining the hamiltonian that maximizes the initial acceleration of a given state, and then posing (and solving analytically) the \emph{brachistophase} problem: for a given initial state, find the (time-independent) hamiltonian that maximizes the  geometric phase accumulated after a given time $\tau$. 

The paper is organized as follows: In Sec.~\ref{TeCPn2un} we review pertinent geometrical aspects of  quantum state space. Covariant derivatives  of a general curve in quantum state space are studied in Sec.~\ref{OtaociP}, including the particular case of a Schr\"odinger curve, \ie, a curve that evolves according to the homonymous equation, with a time-independent hamiltonian. Section~\ref{AagpociP} discusses the relation between the geometric phase and the covariant derivatives of the curve. The maximization problems mentioned above  are studied in Sec.~\ref{Brachi}.  A summary of our results and some concluding comments are presented in Section~\ref{Conclusions}.


\section{Mathematical preliminaries}
\label{TeCPn2un}
\subsection{Coordinates, metric, connection, and curvature of the projective space}
Let $\mathcal{H} \equiv \mathbb{C}^{n+1}$ be the Hilbert space of a spin-$s$ quantum system, where $n=2s$. The elements $\ket{\psi} \in \mathcal{H}$ that differ by a non-zero scalar factor, $\ket{\psi'}=\lambda \ket{\psi}$, with $\lambda \in \mathbb{C}-\{0\}$ form an equivalence class
$\pket{\psi}$. The ket $\ket{\psi} \in \mathcal{H}$ is sent by the projection $\pi$ to  $\pket{\psi}$, the latter being a point in the complex projective space 
$\mathbb{C}P^{n}$, \ie, the space  of complex lines through the origin in $\mathcal{H}$, 
\begin{equation}
	\label{piproj}
	\pi \colon \mathcal{H} \rightarrow \mathbb{C}P^n, \quad \ket{\psi}=(\psi^0,\psi^1,\ldots,\psi^n)^T \mapsto \pket{\psi}=(z^1,\ldots,z^n)
	\, ,
\end{equation}
with the quantities $z^i=\psi^i/\psi^0$, together with their complex conjugates $\bar{z^i} \equiv w^i$ being coordinates  in the  chart $U_0$ of $\mathbb{C}P^n$, where $\psi^0 \neq 0$ --- we will denote them collectively by $z^A$, with $A$ ranging over $\{1,\ldots,n,\bar{1},\ldots,\bar{n}\}$, implying the slight abuse of notation $z^{\bar{i}} \equiv \bar{z^i} \equiv w^i$. 

We denote the group of unitary matrices of dimension $n+1$ and its corresponding Lie algebra of hermitian matrices  by $U(n+1)$ and $\mathfrak{u}(n+1)$, respectively (we follow the physicists' convention in which the structure constants are pure imaginary). $\mathbb{C}P^n$ may be embedded into $\mathfrak{u}(n+1)$ as the $U(n+1)-$adjoint orbit of the density matrix $\rho_0=\text{diag}(1,0,\ldots,0)$ (see, \eg,~\cite{Ban.Hur:04}), the latter living naturally in $\mathfrak{u}(n+1) \cong \mathbb{R}^{(n+1)^2}$,
\begin{equation}
	\label{CPn2un}
	\varphi \colon \mathbb{C}P^n \rightarrow \mathbb{P} \hookrightarrow \mathfrak{u}(n+1) \, ,
	\quad
	\pket{\psi} \mapsto \rho_\psi=\frac{\ket{\psi}\bra{\psi}}{\braket{\psi}{\psi}}=
	\Delta^{-1}
	\left(
	\begin{array}{cccc}
		1 & w^1 & \ldots & w^n
		\\
		z^1 & z^1 w^1 & \ldots & z^1 w^n
		\\
		\vdots & & & \vdots
		\\
		z^n & z^n w^1 & \ldots & z^n w^n
	\end{array}
	\right)
	\, ,
\end{equation}
with $\Delta \equiv 1+\sum_{i=1}^n z^i w^i$.
We denote the image of $\mathbb{C}P^n$ under $\varphi$ by $\mathbb{P} \subset \mathfrak{u}(n+1)$. The dimension of $\rho$ is that of $\mathcal{H}$, equal to $n+1$ --- we enumerate the components of $\ket{\psi}$ and the rows and columns of $\rho$  by greek indices ranging from 0 to $n$. We also use the notation $(z^{\mu})=(1,z^1,\ldots,z^n)$, so that, \eg,
\begin{equation}
	\label{rhomunu}
	\rho^{\mu \nu}=\Delta^{-1} z^\mu w^\nu
	\, .
\end{equation}
A ``basis'' in the tangent space $T_\rho \mathbb{P}$ is  given by the matrices $\rho_A \equiv \partial \rho /\partial z^A$, 
with real tangent vectors $v$ constrained to satisfy $v^{\bar{a}}=\overline{v^a}$, where $v^{\bar{a}}$ denotes the component of $v$ along $\partial_{w^a}$, $v=v^a \partial_{z^a}+v^{\bar{a}} \partial_{w^a} \equiv v^a \partial_a +v^{\bar{a}} \partial_{\bar{a}}$. Note that the matrices $\rho_A$ are not hermitean, and, hence, are not by themselves tangent to $\mathbb{P}$ --- to obtain tangent vectors to $\mathbb{P}$ we need to restrict to real  ones, satisfying the above constraint, a true basis in $T_\rho \mathbb{P}$ is then given by, \eg,  $\{ \rho_i+\rho_{\bar{i}}, i(\rho_i -\rho_{\bar{i}})\}$). In the ``basis'' $\{\rho_A\}$, the Fubini-Study (FS) metric and its inverse have components 
\begin{equation}
	\label{FSmcomp}
	g_{a\bar{b}}=\frac{1}{2}\Delta^{-2}(\Delta\delta^a_{\phantom{a}b}- z^b w^a)
	\, ,
	\quad
	g^{a\bar{b}}=2\Delta(\delta^a_{\phantom{a}b}+z^a w^b)
	\, ,
\end{equation}
with $g_{\bar{b}a}=g_{a\bar{b}}$ (\ie, $g_{AB}$ is symmetric), and $g_{b\bar{a}}=\bar{g}_{a\bar{b}}$ (\ie, $(g_{a\bar{b}})$ is hermitean) and similar statements holding true for the inverse metric. Note that the fact that $g$ comes from a K\"ahler potential ($K=2\log \Delta$) implies that $g_{a\bar{b},c}=g_{c\bar{b},a}$ and $g_{a\bar{b},\bar{c}}=g_{a\bar{c},\bar{b}}$.

The Christoffel symbols are found to be~\cite{Mor:07}
\begin{equation}
	\label{ChrComp}
	\Gamma^{c}_{\phantom{c}ab}
	=
	g^{c\bar{r}} \partial_a g_{b\bar{r}}
	=
	-\Delta^{-1}( \delta^c_{\phantom{c}b} w^a+\delta^c_{\phantom{c}a} w^b)
	\, ,
	\quad
	\Gamma^{\bar{c}}_{\phantom{\bar{c}}\bar{a}\bar{b}}=-\Delta^{-1}( \delta^c_{\phantom{c}b} z^a+\delta^c_{\phantom{c}a} z^b)
	\, ,
\end{equation}
with all mixed components vanishing, while the Riemann tensor is given by
\begin{equation}
	\label{Rcomp}
	R_{a\bar{b}c\bar{d}}=\frac{1}{2}(g_{a\bar{b}} g_{c\bar{d}}+g_{a\bar{d}} g_{c\bar{b}})
	\, ,
\end{equation}
all other independent components being zero~\cite{Mor:07}. Accordingly,
\begin{equation}
	\label{covderrho}
	\nabla^{(g)}_b \rho_a=-\Delta^{-1}(w^a \rho_b+w^b\rho_a)
	\, ,
	\quad
	\nabla^{(g)}_{\bar{b}} \rho_a=0
	\, ,
	\quad
	\nabla^{(g)}_{b} \rho_{\bar{a}}=0
	\, ,
	\quad
	\nabla^{(g)}_{\bar{b}}\rho_{\bar{a}}
	=
	-\Delta^{-1}(z^a \rho_{\bar{b}}+z^b\rho_{\bar{a}})
	\, .
\end{equation}
\subsection{Geometry of the embedding $\mathbb{C}P^n \hookrightarrow \mathfrak{u}(n+1)$}
\label{Gote}
The tangent space $T_{\ket{\psi}}\mathcal{H}$ can be decomposed into parallel and normal subspaces, $T_{\ket{\psi}} \mathcal{H}=P_{\ket{\psi}} \mathcal{H} \oplus N_{\ket{\psi}} \mathcal{H}$, with corresponding projectors $\rho$ and $\tilde{\rho} \equiv I-\rho$. $N_{\ket{\psi}} \mathcal{H}=\tilde{\rho} \, T_{\ket{\psi}} \mathcal{H}$ is isomorphic to $T_{\rho_\psi}\mathbb{P}$ via 
\begin{equation}
	\label{pistarN}
	\pi_*|_{N} \colon N_{\ket{\psi}} \mathcal{H} \ni \ket{v} \mapsto \ket{v}\bra{\psi}+\ket{\psi}\bra{v}
	\equiv X_{\ket{v}}
	\, .
\end{equation}
The FS metric $g$ on $\mathbb{P}$ is obtained from the hermitean inner product in $\mathcal{H}$ via $g(X_{\ket{v}},X_{\ket{u}})=\text{Re} \braket{v}{u}$. The complex structure on $\mathcal{H}$ given by $J(\ket{v})=i \ket{v}$ induces a complex structure (also denoted by $J$) on $\mathbb{P}$ given by 
\begin{equation}
	\label{JPdef}
	J(X_{\ket{v}})=X_{J(\ket{v})}=X_{i\ket{v}}=i \left( \ket{v}\bra{\psi}-\ket{\psi}\bra{v} \right) \equiv \tilde{X}_{\ket{v}}
	\, .
\end{equation}
Note that $i[X_{\ket{v}},\rho]=\tilde{X}_{\ket{v}}$, so that
\begin{equation}
	\label{Jcommrho}
	J(\cdot)=i [\cdot,\rho]
\end{equation}
and, \eg, $X_{\ket{v}}=-J(\tilde{X}_{\ket{v}}) =i[\rho,\tilde{X}_{\ket{v}}]=[[X_{\ket{v}},\rho],\rho]$. An arbitrary matrix $A \in \mathfrak{u}$, considered as hamiltonian operating on $\mathcal{H}$, generates the Schr\"odinger vector field $\ket{\dot{\psi}}=-i A \ket{\psi}$, which projects to the fundamental field $\hat{A}$ on $\mathbb{P}$,
\begin{align}
	\hat{A}
	&=
	\pi_* \ket{\dot{\psi}}
	\nonumber
	\\
	&=
	X_{-i \tilde{\rho} A\ket{\psi}}
	\nonumber
	\\
	&=
	-i (I-\rho) A\ket{\psi} \bra{\psi}+i \ket{\psi}\bra{\psi}A (I-\rho)
	\nonumber
	\\
	&=
	-i A \rho +i \rho A \rho + i \rho A -i \rho A \rho
	\nonumber
	\\
	&=
	i[\rho,A]
	\, .
	\label{hatA}
\end{align} 

The natural metric in $\mathfrak{u}$ is given by $G(X,Y)=\frac{1}{2} \text{Tr}(XY)$, which is invariant under the adjoint action of $U(n+1)$,
\begin{equation}
	\label{natmetr}
	G(\text{Ad}_g(X),\text{Ad}_g(Y))=G(X,Y)
	\, ,
\end{equation}
the infinitesimal version of which is 
\begin{equation}
	\label{invGinf}
	G([Z,X],Y)+G(X,[Z,Y])=0
	\, .
\end{equation}
The tangent space $T_\rho \mathfrak{u}$, $\rho \in \mathbb{P}$,  can be decomposed in subspaces tangent and normal to $\mathbb{P}$ respectively (in the metric  $G$), $T_\rho \mathfrak{u}=T_\rho \mathbb{P} \oplus N_\rho \mathbb{P}$, with $T_{\rho}\mathbb{P} \perp N_{\rho}\mathbb{P}$.  The vectors $\hat{A}$, with $A \in \mathfrak{u}$, generate $T_\rho \mathbb{P}$, since the action of $U(n+1)$ on $\mathbb{P}$ is transitive --- symbolically, 
\begin{equation}
	\label{TrhoPgen}
	T_\rho \mathbb{P}=i[\rho,\mathfrak{u}]
	\, .
\end{equation}
Note that both $A$ and $\hat{A}$ are matrices in $\mathfrak{u}$, and if $A$ belongs to $T_\rho \mathbb{P}$ then $\hat{A}=J(A)$ also belongs to $T_\rho \mathbb{P}$, and vice versa. 
As shown in~\cite{Ban.Hur:04}, the normal space $N_\rho \mathbb{P}$ is generated by matrices $B \in \mathfrak{u}$ such that $[\rho,B]=0$. In view of~(\ref{hatA}), this means that the normal part, \wrt{} $\rho$, of an $A \in \mathfrak{u}$, does not contribute to $\hat{A}(\rho)$. For a given $D \in \mathfrak{u}$ and $\rho \in \mathbb{P}$, we define the even and odd part of $D$ (\wrt{} to $\rho$)  by
\begin{equation}
	\label{Deodef}
	D^{\text{e}}=\rho D \rho +\tilde{\rho} D \tilde{\rho}
	\, ,
	\qquad
	D^{\text{o}}=\rho D \tilde{\rho} +\tilde{\rho} D \rho
	\, ,
\end{equation}
and $D=D^{\text{e}}+D^{\text{o}}$. 
It can be shown that $\text{ad}_{D^{\text{e}}}$ maps  $T_\rho \mathbb{P}$ to $T_\rho \mathbb{P}$, and $N_\rho \mathbb{P}$ to $N_\rho \mathbb{P}$, while $\text{ad}_{D^{\text{o}}}$ maps $T_\rho \mathbb{P}$ to $N_\rho \mathbb{P}$ and \emph{vice-versa}. Indeed, for a tangent vector $[\rho,X]$, we have
\begin{equation}
	\label{DeTT}
	[D^{\text{e}},[\rho,X]]=[\rho,[D^{\text{e}},X]]-[[\rho,D^{\text{e}}],X]=[\rho,[D^{\text{e}},X]]
	\, ,
\end{equation}
where, in the first equality we used the Jacobi identity, while in the second one the fact that $[\rho,D^{\text{e}}]=0$. Thus, $\text{ad}_{D^{\text{e}}}$ sends $[\rho,X] \in T_{\rho}\mathbb{P}$ to $[\rho,[D^{\text{e}},X]] \in T_{\rho}\mathbb{P}$. On the other hand, for a tangent vector $X_{\ket{v}}=\ket{v}\bra{\psi}+\ket{\psi}\bra{v}$ (with $\braket{v}{\psi}=0$), we have
\begin{equation*}
	[D^{\text{o}},X_{\ket{v}}]=\rho D \ket{v}\bra{\psi}+\tilde{\rho} D \ket{\psi}\bra{v}-\ket{v}\bra{\psi}D \tilde{\rho}-\ket{\psi}\bra{v}D\rho
	\, ,
\end{equation*}
from which it easily follows that $[\rho,[D^{\text{o}},X_{\ket{v}}]]=0$, implying that $[D^{\text{o}},X_{\ket{v}}] \in N_{\rho}\mathbb{P}$. Similarly, for $Z \in N_{\rho}\mathbb{P}$, \ie, such that $[\rho,Z]=0$, we have $[\rho,[D^{\text{e}},Z]]=[[\rho,D^{\text{e}}],Z]+[D^{\text{e}},[\rho,Z]]=0$, so that $[D^{\text{e}},Z] \in N_{\rho}\mathbb{P}$. Also, for $W \in N_{\rho}\mathbb{P}$, 
\begin{align*}
	\text{Tr} \left( W [D^{\text{o}},Z]] \right)
	&=
	\text{Tr} \left( W \rho D \tilde{\rho}Z+W\tilde{\rho}D\rho Z-WZ \rho D \tilde{\rho}-WZ \tilde{\rho} D \rho \right)
	\\
	&=
	\text{Tr} \left( \rho WDZ \tilde{\rho}+\tilde{\rho} WDZ \rho-\rho WZD \tilde{\rho} -\tilde{\rho} WZD \rho \right)
	\\
	&=
	0
	\, ,
\end{align*}
where the fact that $Z$, $W$ commute with $\rho$, $\tilde{\rho}$ was used. Thus, $[D^{\text{o}},Z]$ is orthogonal to $N_{\rho}\mathbb{P}$, and, hence, it belongs to $T_{\rho}\mathbb{P}$. 

It can also be seen easily that the above decomposition of a general hermitean matrix $D$, regarded as a tangent vector to $\mathfrak{u}$ at $\rho$, coincides with the decomposition $T_{\rho}\mathfrak{u}=T_{\rho}\mathbb{P} \oplus N_{\rho}\mathbb{P}$, with $D^{\text{o}} \in T_{\rho}\mathbb{P}$ and $D^{\text{e}} \in N_{\rho}\mathbb{P}$. Indeed, $[\rho,D^{\text{e}}]=0$ implies that $D^{\text{e}} \in N_{\rho}\mathbb{P}$, while for any $Z \in N_{\rho}\mathbb{P}$, 
\begin{align*}
	\Tr \left( D^{\text{o}} Z \right)
	&=
	\Tr \left(
	\rho D \tilde{\rho} Z+\tilde{\rho}D \rho Z \right)
	\\
	&=
	\Tr \left(
	\rho DZ\tilde{\rho} +\tilde{\rho}DZ\rho \right)
	\\
	&=
	0
	\, ,
\end{align*}
so that $D^{\text{o}} \perp N_{\rho}\mathbb{P}$, implying that $D^{\text{o}} \in T_{\rho}\mathbb{P}$. Note that 
\begin{equation}
	\label{oddprojcomm}
	D^{\text{o}}=\rho D \tilde{\rho} +\tilde{\rho} D \rho =[\rho,[\rho,D]]
	\, ,
\end{equation}
\ie, projection onto the tangent space of $\mathbb{P}$ is obtained by a double commutator with $\rho$. Another way to obtain the last result is by writing $D=D^{\text{e}} + D^{\text{o}}$, and noting that the normal (even) part is filtered out in the first commutator, $[\rho,D]=[\rho,D^{\text{o}}]=i J(D^{\text{o}})$ so that the double commutator just gives $-J^2(D^{\text{o}})=D^{\text{o}}$.
\section{On the acceleration of curves in $\mathbb{P}$}
\label{OtaociP}
\subsection{General curves in $\mathbb{P}$}
\label{GciP}
Consider any basis $\{E_{\mu \nu} \}$ of $T\mathfrak{u}$, with position-independent entries $(E_{\mu\nu})_{\sigma\tau}$ (we choose to enumerate the $(n+1)^2$ elements of the basis by the composite index $\mu \nu$). The metric $G$ on $\mathfrak{u}$, in that basis, has position-independent components,  so that the Christoffel symbols vanish, and, for a position-dependent matrix $A$, interpreted as a vector field on $\mathfrak{u}$, we have
\begin{equation}
	\label{covderXA}
	(\nabla^{(G)}_X A)_{\mu \nu}=X \triangleright A_{\mu \nu}
	\, .
\end{equation}
In the vicinity of $\mathbb{P}$ in $\mathfrak{u}$ one may choose coordinates for $\mathfrak{u}$ complementing the $2n$ coordinates $z^A$ on $\mathbb{P}$ by $n^2-2n$ additional coordinates, transversal to $\mathbb{P}$, and extending $z^A$ continuously in a neighborhood of $\mathbb{P}$. Then,  $\nabla^{(G)}_a \rho_b=\partial_a \rho_b\equiv \rho_{ab}$ and a short(ish) calculation, starting from~(\ref{rhomunu}),  shows that 
\begin{align}
	(\rho_a)^{\mu \nu} 
	&=
	\Delta^{-2} \left( \Delta \delta^\mu_{\phantom{\mu} a} w^\nu-z^\mu w^\nu w^a \right)
	\, ,
	\label{rhoa}
	\\
	(\rho_{\bar{b}})^{\mu \nu}
	&=
	\Delta^{-2} \left( \Delta \delta^{\nu}_{\phantom{\nu}b} z^\mu-z^b z^\mu w^\nu \right)
	\, ,
	\label{rhobb}
	\\
	(\rho_{ab})^{\mu \nu} 
	&=
	\Delta^{-3} \left(
	2 z^\mu w^\nu w^a w^b -\Delta 
	\left(
	\delta^\mu_{\phantom{\mu}a} w^b w^\nu+\delta^\mu_{\phantom{\mu} b} w^a w^\nu
	\right)
	\right)
	\, ,
	\label{rhoab}
	\\
	(\rho_{a\bar{b}})^{\mu \nu} 
	&=
	\Delta^{-3} \left(
	2z^b z^\mu w^a w^\nu -\Delta 
	\left(
	\delta^\mu_{\phantom{\mu}a}z^b w^\nu
	+
	\delta^{\nu}_{\phantom{\nu}b} z^\mu w^a
	+
	\delta^a_{\phantom{a}b} z^\mu w^\nu
	-
	\Delta \delta^\mu_{\phantom{\mu}a} \delta^\nu_{\phantom{\nu} b}
	\right)
	\right)
	\, ,
	\\
	(\rho_{\bar{a}\bar{b}})^{\mu \nu} 
	&=
	\Delta^{-3} \left(
	2z^a z^b z^\mu w^\nu -\Delta 
	\left(
	\delta^\nu_{\phantom{\mu}a}z^b z^\mu
	+
	\delta^{\nu}_{\phantom{\nu}b} z^a z^\mu 
	\right)
	\right)
	\, .
	\label{rhoabb}
\end{align}
Finally, we use~(\ref{oddprojcomm}), with a change in notation for $D^{\text{o}} \rightarrow D^\parallel$ and $D^{\text{e}} \rightarrow D^\perp$, to obtain
\begin{align}
	\rho_{ab}^\parallel
	&=
	-\Delta^{-1} \left(w^a \rho_b+w^b \rho_a \right)
	\, ,
	\label{rhoabpar}
	\\
	\rho_{a\bar{b}}^\parallel
	&=
	0
	\, ,
	\label{rhoabbpar}
	\\
	\rho_{\bar{a}\bar{b}}^\parallel
	&=
	-\Delta^{-1} \left( z^a \rho_{\bar{b}}+z^b \rho_{\bar{a}} \right)
	\, ,
	\label{rhobabbpar}
\end{align}
implying that (compare~(\ref{rhoabpar}), (\ref{rhoabbpar}), (\ref{rhobabbpar}) to~(\ref{covderrho})), for $X$, $Y \in T_\rho \mathbb{P}$,  $\nabla^{(g)}_X Y=(\nabla^{(G)}_X Y)^\parallel$. In other words, the Levi-Civita covariant derivative  for $g$ on $\mathbb{P}$ may be obtained by projecting the ambient covariant derivative (corresponding to the euclidean $G$ on $\mathfrak{u}$) onto the tangent space of $\mathbb{P}$, as in the standard treatment of \eg,  surfaces in $\mathbb{R}^3$. 

Given a curve $\rho_t$ in $\mathbb{P}$, its velocity is $v=\dot{\rho}_t \in T_{\rho_t}\mathbb{P}$ while $\dot{v}$ (when $\rho_t$ is viewed as a curve in $\mathfrak{u}$) is $\dot{v}=\nabla^{(G)}_t v=\ddot{\rho}_t$, so that (dropping the subscript $t$)
\begin{align}
	\alpha \equiv \nabla^{(g)}_t v & = \nabla^{(g)}_t \dot{\rho}
	\nonumber
	\\
	&=
	\left( \nabla_t^{(G)} \dot{\rho} \right)^\parallel
	\nonumber
	\\
	&=
	\ddot{\rho}^{\, \parallel}
	\nonumber
	\\
	&=
	\rho \, \ddot{\rho} \tilde{\rho} +\tilde{\rho} \, \ddot{\rho} \rho
	\, .
	\label{alphares}
\end{align}

Note that for any $A \in T_\rho \mathfrak{u}$, with $\rho \in \mathbb{P}$,
\begin{equation}
	|A^\parallel|^2
	=
	\frac{1}{2} \text{Tr} \left( \left(A^\parallel \right)^2 \right)
	=
	\frac{1}{2} \text{Tr} 
	\left( 
	\left( \rho A \tilde{\rho} +\tilde{\rho} A \rho \right)^2 
	\right)
	=
	\, 
	< \! A^2 \! >-< \! A \! >^2
	\, ,
	\label{Aparsq}
\end{equation}
so that 
\begin{equation}
	\label{vdmsq}
	|\alpha|^2=< \! \ddot{\rho}^{\, 2} \! >-< \! \ddot{\rho} \! >^2
	\, . 
\end{equation}

It might prove useful to cast~(\ref{alphares}) in a ``big matrix'' form. An arbitrary $A \in \mathfrak{u}$ may be represented by a $n^2$-dimensional vector $\ket{A}$ containing the entries $A_{\mu \nu}$ in the standard order, $\ket{A}=(A_{11},A_{12},\ldots,A_{nn})^T$. Then (\ref{alphares}) can be written as $\ket{\alpha}=\mathbf{P} \ket{\ddot{\rho}}$, with the $n^2 \times n^2$ hermitean matrix $\mathbf{P}$ given by 
\begin{equation}
	\label{bfP}
	\mathbf{P}^{\mu \nu,\alpha\beta}=\delta^\beta_{\phantom{\beta}\nu} \rho^{\mu \alpha} +\delta^\mu_{\phantom{\mu}\alpha} \rho^{\beta \nu} -2 \rho^{\mu \alpha} \rho^{\beta\nu}
	\, ,
\end{equation}
or, in matrix tensor product form,
\begin{align}
	\label{bfPtpf1}
	\mathbf{P}
	&=
	\rho \otimes \tilde{\rho}^T+\tilde{\rho} \otimes \rho^T
	\\
	&=
	\rho \otimes I + I \otimes \rho^T -2 \rho \otimes \rho^T
	\, ,
	\label{bfPtpf2}
\end{align}
with $(A \otimes B^T) \ket{C}=\ket{ACB}$.
In this notation, $\braket{A}{B}=\bar{A}_{\mu \nu} B_{\mu \nu}=A_{\nu \mu}B_{\mu \nu}=\text{Tr}(AB)=2G(A,B)$. 
Being a projection operator, $\mathbf{P}$ satisfies $\mathbf{P}^2=\mathbf{P}$. Since projection onto the tangent space of $\mathbb{P}$ is effected by a double commutator with $\rho$, we get $\mathbf{P}=\mathfrak{R}^2$, where $\mathfrak{R}=\rho \otimes I - I \otimes \rho^T$ is the adjoint representation of $\rho$, with $\mathfrak{R}\ket{A}=\ket{[\rho,A]}$, so that
\begin{align}
	\label{alphabmf}
	\alpha
	&=
	[\rho,[\rho,\ddot{\rho}]]
	\\
	\ket{\alpha} 
	&=
	\mathfrak{R}^2 \ket{\ddot{\rho}}
	\, .
\end{align}
For the modulus squared of $\alpha$ we find
\begin{equation}
	\label{}
	|\alpha|^2=G(\alpha,\alpha)
	=
	\frac{1}{2} \text{Tr}(\alpha^2)
	=
	\frac{1}{2} \braket{\alpha}{\alpha}
	=
	\frac{1}{2}\bra{\ddot{\rho}}\mathbf{P} \ket{\ddot{\rho}}
	=
	\frac{1}{2}\bra{\ddot{\rho}}\mathfrak{R}^2 \ket{\ddot{\rho}}
	=
	\frac{1}{2} |\mathfrak{R} \ket{\ddot{\rho}}|^2
	\, ,
\end{equation}
where $\mathbf{P}^\dagger=\mathbf{P}$  and $\mathbf{P}^2=\mathbf{P}$ were used. 
\subsection{Schr\"odinger curves in $\mathbb{P}$}
\label{SciP}
By Schr\"odinger curves in $\mathbb{P}$ we mean curves $\rho_t$ that are solutions to the Schr\"odinger equation, $\dot{\rho}=-i [H,\rho]$. We will limit our attention to the case where $H$ does not depend on time, then $\rho_t=e^{-i t H} \rho_0 e^{i t H}$ and 
\begin{equation}
	\label{ddrH}
	\ddot{\rho}=-[H, [H,\rho]]=2H\rho H-\rho H^2 -H^2 \rho
	\, .
\end{equation}

We compute
\begin{align}
	\ddot{\rho}^2
	&=
	(2H\rho H-\rho H^2 -H^2 \rho)^2
	\nonumber
	\\
	&=
	4 H \rho H^2 \rho H + \rho H^2 \rho H^2+H^2 \rho H^2 \rho 
	-2 H \rho H \rho H^2 -2 \rho H^3 \rho H
	\nonumber
	\\
	& \qquad -2 H \rho H^3 \rho -2 H^2 \rho H \rho H + \rho H^4 \rho + H^2 \rho H^2
	\, ,
\end{align}
so that
\begin{equation}
	\label{ddrev}
	< \! \ddot{\rho}^{\, 2} \! >= \ldots = h_4 - 4 h_1 h_3+3 h_2^2
	\, ,
	\qquad
	< \! \ddot{\rho} \! >^2= 4(h_1^2-h_2)^2=4h_2^2  -8 h_2 h_1^2 +4h_1^4
	\, ,
\end{equation}
and, finally, from~(\ref{vdmsq}) we get
\begin{equation}
	\label{alphasqexpl}
	|\alpha|^2=\ldots =h_4-4h_3h_1 -h_2^2 +8 h_2 h_1^2 -4 h_1^4
	\, ,
\end{equation}
where $h_m \equiv \bra{\psi} H^m \ket{\psi}$.

In the ``big matrix'' notation, $\ket{\ddot{\rho}}=-\mathfrak{H}^2 \ket{\rho}$, where $\mathfrak{H}$ is the adjoint representation of $H \in \mathfrak{u}$ (see the first of~(\ref{ddrH})), $\mathfrak{H}=H \otimes I-I \otimes H^T$, so that
\begin{align}
	\label{alphabmfH}
	\alpha
	&=
	-[\rho,[\rho,[H,[H,\rho]]]]
	\\
	\ket{\alpha} 
	&=
	-\mathfrak{R}^2 \mathfrak{H}^2 \ket{\rho}
	\, .
\end{align}
For the modulus squared of $\alpha$ we find
\begin{equation}
	\label{alphamodsq2}
	|\alpha|^2=
	\frac{1}{2}\bra{\rho}\mathfrak{H}^2 \mathbf{P} \mathfrak{H}^2 \ket{\rho}
	=
	\frac{1}{2} \bra{\rho} \mathfrak{H}^2 \mathfrak{R}^2 \mathfrak{H}^2 \ket{\rho}
	\, ,
\end{equation}
where $\mathfrak{H}^\dagger=\mathfrak{H}$ was used. Introducing the ``density matrix of the density matrix'' $R=\ket{\rho}\bra{\rho}=\rho \otimes \rho^T$, the above can also be written as
\begin{equation}
	\label{ams}
	|\alpha|^2=\frac{1}{2} \text{Tr} \left( R \, \mathfrak{H}^2 \mathfrak{R}^2 \mathfrak{H}^2 \right)
	\, .
\end{equation}
Finally, we may also define the acceleration $a$ of the curve $\rho_t$ as its second covariant derivative \wrt{} length, rather than time. Since the modulus of the velocity $v$ of a curve $\rho_t$, describing time evolution generated by Schr\"odinger's equation with a time-independent hamiltonian $H$ is constant in time, $\partial_t |v|=0$, we get $a=\alpha/|\dot{s}|$, and 
\begin{equation}
	\label{alphavdot}
	|a|^2=\dot{s}^{-2} |\alpha|^2/=(h_2-h_1^2)^{-1}|\alpha|^2
	\, .
\end{equation}
Note that, in this case, the modulus of $a$ is the curvature of the curve. 

\section{Geometric phase and covariant derivatives of curves in $\mathbb{P}$}
\label{AagpociP}
A generalization of the standard geometric phase, valid for open (\ie, non-cyclic) curves,
is given in~\cite{Muk.Sim:93}. As shown there, the  geometric phase accumulated along a geodesic (in the Fubini-Study metric) is zero.
On the other hand, geodesics are characterized by their vanishing acceleration.
It seems reasonable then to inquire about the relation between the acceleration of a curve and the associated
geometric phase --- note that this relation ought to exist independently of the Schr\"odinger dynamics.

Given a curve $\rho_{\tau}$ , $0 \leq \tau \leq T$, in $\mathbb P$, the accumulated geometric phase up to a time $0 \leq t\leq T$
is ~\cite{Muk.Sim:93},
\begin{equation}
	\label{gpMS}
	\varphi_t=\arg \Tr \left( \rho_0 F_t \right)
	\, ,
\end{equation}
where
\begin{equation}
	\label{Ftdef}
	F_t=\mathcal{P}\!\exp\int_0^t \td \tau \dot{\rho}_{\tau}
	\, ,
\end{equation}
and a dot denotes a time derivative, while $\mathcal{P}\!\exp$ a path-ordered exponential.
By definition, this means that $F_\tau$ satisfies the differential equation
\begin{equation}
	\label{Ftde}
	\dot{F}_\tau=\dot{\rho}_\tau F_\tau
	\, ,
\end{equation}
subject to the initial condition $F_0=I$. The open-curve phase
$\varphi_t$ has a simple geometrical interpretation: it is the usual Berry phase of the closed curve obtained by gluing the curve 
$\rho_{\tau}$, $0 \leq \tau \leq t$ to the geodesic that connects $\rho_t$ with $\rho_0$. In what follows, we find an explicit formula for the derivatives of $\varphi_{t}$ at $t=0$
in terms of $\rho_\tau$ quantities intrinsic to $\Ps$.

\subsection{Explicit computation of derivatives of the geometric phase at $t=0$}
\label{Ecodgp}
\subsubsection{The first three derivatives of the geometric phase}
\label{Tftd}
Using the notation $\evc{A}_t \equiv \Tr \left( \rho_0 A F_t \right)$ for an arbitrary time-dependent operator $A$,
one gets $\partial_t \evc{A}_t=\evc{\dot{A}+A\dot{\rho}}_t$, so that, \eg, (dropping the index $t$)
\begin{equation}
	\label{ptid}
	\partial_t \evc{I} =\evc{ \dot{\rho}}
	\, ,
	\qquad
	\partial_t^2 \evc{I} =\evc{\ddot{\rho} +\dot{\rho}^2}
	\, ,
	\qquad
	\partial_t^3 \evc{ I } =\evc{ \dddot{\rho} + 2 \ddot{\rho} \dot{\rho} + \dot{\rho} \ddot{\rho} +\dot{\rho}^3 }
	\, .
\end{equation}
Note that, by (\ref{gpMS}), $\varphi_t=\mathrm{Im} \log \evc{I}_t$, so that
\begin{align}
	\dot{\varphi}
	&=
	\mathrm{Im} \left( \evc{I}^{-1} \evc{\dot{\rho}} \right)
	\\
	\ddot{\varphi}
	&=
	\mathrm{Im} \left( \evc{I}^{-2} \left(
	\evc{I} \evc{\ddot{\rho} +\dot{\rho}^2} - \evc{\dot{\rho}}^2 \right) \right)
	\\
	\dddot{\varphi}
	&=
	\mathrm{Im} \left(\evc{I}^{-3} \left(
	\evc{I}^2 \evc{\dddot{\rho} +2\ddot{\rho} \dot{\rho} +\dot{\rho} \ddot{\rho} +\dot{\rho}^3}
	-3 \evc{I} \evc{ \dot{\rho}} \evc{\ddot{\rho}+\dot{\rho}^2}
	+2 \evc{\dot{\rho}}^3
	\right)\right)
	\, .
\end{align}
At $t=0$, $\evc{I}_0=\Tr(\rho_0)=1$ and $\evc{\dot{\rho}}_0=\Tr(\rho_0 \dot{\rho}_0)=0$, resulting in
\begin{align}
	\dot{\varphi}_0
	&=
	0
	\\
	\ddot{\varphi}_0
	&=
	\mathrm{Im}\,
	\evc{ \ddot{\rho}+\dot{\rho}^2}_0
	\\
	\dddot{\varphi}_0
	&=
	\mathrm{Im} \,
	\evc{\dddot{\rho} +3\dot{\rho} \ddot{\rho}+\dot{\rho}^3}_0
	\, .
\end{align}
All time derivatives of $\rho$ are hermitean so, using cyclicity of the trace gives
\begin{align}
	\ddot{\varphi}_0
	&=
	\mathrm{Im} \,
	\Tr \left( \rho_0 \ddot{\rho}_0 +\rho_0 \dot{\rho}_0^2 \right)
	\nonumber
	\\
	&=
	\frac{1}{2i} \left( 
	\Tr \left( 
	\rho_0 \ddot{\rho}_0 + \rho_0 \dot{\rho}_0^2 \right)
	- \Tr \left( 
	\ddot{\rho}_0 \rho_0  + \dot{\rho}_0^2 \rho_0 \right) 
	\right)
	\nonumber
	\\
	&=
	0
	\label{varphidd0}
	\, ,
\end{align}
leading to the conclusion that  the first two time derivatives of $\varphi_t$ vanish, at $t=0$, for \emph{every} curve $\rho_t$.
Similarly,
\begin{align}
	\dddot{\varphi}_0
	&=
	\mathrm{Im} \,
	\Tr \left( \rho_0 (\dddot{\rho}_0 + 2 \ddot{\rho}_0 \dot{\rho}_0 + \dot{\rho}_0\ddot{\rho}_0 +\dot{\rho}_0^3) \right)
	\nonumber
	\\
	&=
	\frac{1}{2i} \left( 
	\Tr \left( 
	\rho_0 \dddot{\rho}_0 + 2 \rho_0 \ddot{\rho}_0 \dot{\rho}_0 + \rho_0 \dot{\rho}_0\ddot{\rho}_0+ \rho_0 \dot{\rho}^3_0 
	\right)
	- \Tr \left( 
	\dddot{\rho}_0 \rho_0  + 2 \dot{\rho}_0  \ddot{\rho}_0 \rho_0 +  \ddot{\rho}_0 \dot{\rho}_0 \rho_0 +\dot{\rho}_0^3 \rho_0
	\right) 
	\right)
	\nonumber
	\\
	&=
	\frac{1}{2i}  \Tr \left( \rho_0 [\ddot{\rho}_0,\dot{\rho}_0] \right)
	\nonumber
	\\
	&=
	\frac{1}{2i}  \Tr \left( \dot \rho_0 [\rho_0,\ddot {\rho}_0] \right)
	\, .
	\label{varphiddd0}
\end{align}
The quantity in the \rhs{} of~(\ref{varphiddd0}) is, for a general curve $\rho_t$, nonzero, so the first nonzero derivative of $\varphi_{t}$ for a general curve is the third one. The matrix $\ddot{\rho}_0$ in~(\ref{varphiddd0}) represents a vector tangent to the ambient vector space $\mathfrak{u}(n+1)$, but not to $\mathbb{P}$ --- we may remedy this noting that, for a general curve $\rho_t$,
only the part of $\ddot{\rho}_{0}$ tangential to $\mathbb{P}$, $\ddot{\rho}_{0}^\parallel \equiv \alpha$,  contributes to
$\dddot{\varphi}_0$ since
$[\rho_{0},\ddot{\rho}_{0}]=[\rho_{0},\ddot{\rho}_{0}^\parallel]$, so we may write (with $\dot{\rho}_{0}= v$),
\begin{align}
	\dddot{\varphi}_0
	& \, \, =
	\frac{1}{2i} \Tr \left(
	v [\rho_0,\alpha] \right)
	\nonumber
	\\
	&\, \, =
	\frac{1}{2} \Tr \left(
	v J(\alpha)
	\right)
	\nonumber
	\\
	& \, \, =
	g \left( v,J(\alpha) \right)
	\nonumber
	\\
	& \, \, =
	\omega(\alpha,v)
	\, ,
	\label{varphidddav}
\end{align}
where~(\ref{Jcommrho}) was used to obtain the second line,
\ie, $\dddot{\varphi}_0$ is equal to the symplectic area of the parallelogram spanned by the (initial) velocity
and acceleration of the curve. It follows that, for geodesics, where $\alpha=0$, $\dddot{\varphi}_0$ vanishes, a result that we extend below to derivatives of all orders.

For Schr\"odinger curves, a short calculation gives
\begin{align}
	\dddot{\varphi}_0
	&=
	-\frac{1}{2}  \Tr \left(
	\rho_0
	\left[
	\left[H, [H,\rho_0] \right] ,
	[\rho_0,H]
	\right]
	\right)
	= h_3-3h_2h_1 +2h_1^3
	\, ,
	\label{varphiddd0H}
\end{align}
where $h_{i}$, as defined in section \ref{SciP}, is evaluated at $\rho_0$.
\subsubsection{The fourth and fifth derivative of the geometric phase}
\label{Tfd}
For the fourth time derivative of $\varphi$ we find
\begin{equation}
	\label{d4varphi}
	\varphi^{(4)}_0=
	\frac{1}{2i} \Tr \left(
	2 \rho [ \dddot{\rho}^{\, \parallel}, \dot{\rho}]+3 \rho [ \alpha,\dot{\rho}^2] \right)
	\, .
\end{equation}
We proceed to express this in terms of the covariant derivatives of $v$. From $\rho^2=\rho$ one gets
\begin{equation}
	\label{rhorhod}
	\rho \dot{\rho}+\dot{\rho} \rho =\dot{\rho}
	\,,
\end{equation}
which implies
\begin{equation}
	\label{rdreq1}
	\rho \dot{\rho}=\dot{\rho}\tilde{\rho}
	\, ,
	\qquad
	\tilde{\rho} \dot{\rho} =\dot{\rho} \rho
	\, ,
\end{equation}
so that $\rho \dot{\rho}^2=\dot{\rho}^2 \rho$, \ie, $\dot{\rho}^2 \in N_\rho \mathbb{P}$, and the second term in the \rhs{} of~(\ref{d4varphi}) vanishes. Putting $\beta \equiv \nabla^{(g)}_t \alpha= \dot{\alpha}^\parallel$, and using
\begin{equation}
	\label{betarhod3}
	\dddot{\rho}^{\, \parallel} = \beta -[\rho, [\dot{\rho}, \ddot{\rho}] ]
	\, ,
\end{equation}
we find from~(\ref{d4varphi}) (putting $\dot{\rho} \rightarrow v$)
\begin{equation}
	\label{d4varphi2}
	\varphi^{(4)} = -i \Tr \left(
	\rho [\beta,v]
	\right)
	+i \Tr \left(
	\rho [ [\rho, [v,\ddot{\rho}]] , v] \right)
	\, .
\end{equation}
Note that
\begin{align*}
	\Tr \left(
	\rho [ [\rho, [v,\ddot{\rho}]] , v] \right)
	&=
	\Tr \left(
	v [\rho, [ \rho, [ v,\ddot{\rho}]]] \right)
	\\
	&=
	\Tr \left(
	v [ v, \ddot{\rho}]^\parallel \right)
	\\
	&=
	\Tr \left(
	v (\rho v \ddot{\rho} \tilde{\rho} +\tilde{\rho} v \ddot{\rho} \rho
	- \rho \ddot{\rho} v \tilde{\rho} - \tilde{\rho} \ddot{\rho} v \rho)
	\right)
	\\
	&=
	\Tr \left(
	\tilde{\rho} v^2 \ddot{\rho} + \rho v^2 \ddot{\rho} -\tilde{\rho} v \ddot{\rho} v - \rho v \ddot{\rho} v \right)
	\\
	&=
	\Tr \left( v^2 \ddot{\rho} -v \ddot{\rho} v \right)
	\\
	&=
	0
	\, ,
\end{align*}
so that the second term in the \rhs{} of~(\ref{d4varphi2}) vanishes, and we arrive at
\begin{equation}
	\label{d4varphif}
	\varphi^{(4)}_0 = 2 \omega(\beta,v)
	\, .
\end{equation}
Next, we proceed with the fifth derivative.
Denote by $\gamma$ the third covariant derivative of $v$, $\gamma=\nabla_t^{(g)}\beta$. The fifth derivative of $\varphi$ at $t=0$ involves ${\rho^{(4)}}^\parallel_0$, which, in turn, can be expressed in terms of $\gamma$ and lower order $t$-derivatives.
Expressing the latter in terms of $\beta$, $\alpha$, we find
\begin{equation}
	\varphi^{(5)}_0
	=
	\frac{1}{2i} \Tr \left(
	\rho
	\left(
	3[\gamma,v]
	+2[\beta,\alpha]
	-10 \Tr(\rho(\ddot{\rho}+\dot{\rho}^2))[\alpha,v]
	+5[\beta,v^2]
	+5[\ddot{\rho}^2,v]
	+2[\alpha,v^3]
	\right)
	\right)
	\, .
	\label{der5phi}
\end{equation}
The first two terms on the right only involve covariant geometric quantities explicitly --- the rest need some work. From~(\ref{rhorhod}), taking one more derivative, one gets
\begin{equation}
	\label{rhodd1}
	\ddot{\rho}=\rho \ddot{\rho}+\ddot{\rho}\rho+2\dot{\rho}^2
	\, .
\end{equation}
Multiplying by $\rho$ and taking trace one finds
\begin{equation}
	\label{trddrho}
	\Tr(\rho\ddot{\rho})=-2\Tr(\rho\dot{\rho}^2)
	\, ,
\end{equation}
which makes the third term in the \rhs{} of~(\ref{der5phi}) equal to $10 \Tr(\rho\dot{\rho}^2)[\alpha,v]$. On the other hand,
putting $\rho=\ket{\psi}\bra{\psi}$, with $\braket{\dot{\psi}}{\psi}=0$, one gets $\dot{\rho}=\ket{\psi}\bra{\dot{\psi}}+\ket{\dot{\psi}}\bra{\psi}$, so that
\begin{align}
	\dot{\rho}^2
	&=
	\mu \rho+\ket{\dot{\psi}}\bra{\dot{\psi}}
	\label{drho2}
	\\
	\dot{\rho}^3
	&=
	\mu \dot{\rho}
	\label{drho3}
	\, ,
\end{align}
where $\mu \equiv \braket{\dot{\psi}}{\dot{\psi}}$. Note also that, from~(\ref{drho2}), it is easily inferred that
\begin{equation}
	\label{fupro}
	2\Tr(\rho \dot{\rho}^2)=\Tr(\dot{\rho}^2)=2\mu
	\, ,
\end{equation}
so that the third and sixth terms in the \rhs{} of~(\ref{der5phi}) sum to $12\Tr(\rho v^2)[\alpha,v]=6\Tr(v^2)[\alpha,v]=12g(v,v)[\alpha,v]$. The fourth term in the \rhs{} of~(\ref{der5phi}) is zero because $[\rho,v^2]=0$. For the fifth term, start with
\begin{align}
	\ddot{\rho}
	&=
	\ket{\ddot{\psi}}\bra{\psi}+2\ket{\dot{\psi}}\bra{\dot{\psi}}+\ket{\psi}\bra{\ddot{\psi}}
	\\
	\ddot{\rho}^2
	&=
	-\mu \ket{\psi}\bra{\ddot{\psi}}
	+2 \braket{\ddot{\psi}}{\dot{\psi}} \ket{\psi}\bra{\dot{\psi}}
	+\braket{\ddot{\psi}}{\ddot{\psi}} \rho
	\\
	&
	\quad
	{}+ 2\mu \ket{\dot{\psi}} \bra{\dot{\psi}}+2 \braket{\dot{\psi}}{\ddot{\psi}} \ket{\dot{\psi}} \bra{\psi}+
	\ket{\ddot{\psi}} \bra{\ddot{\psi}}-\mu \ket{\ddot{\psi}}\bra{\psi}
\end{align}
where $\braket{\ddot{\psi}}{\psi}=\braket{\psi}{\ddot{\psi}}=-\mu$ has been used (derived by taking derivative of $\braket{\dot{\psi}}{\psi}=0$). A straightforward calculation now shows that $\Tr(\rho[\ddot{\rho}^2,\dot{\rho}])=0$, so that, finally,
\begin{align}
	\varphi^{(5)}_0
	&=
	\frac{1}{2i}\Tr
	\left( \rho \left(
	3[\gamma,v]+2[\beta,\alpha]+12g(v,v)[\alpha,v]
	\right) \right)
	\nonumber
	\\
	&=
	3\omega(\gamma,v)+2\omega(\beta,\alpha)+12  g(v,v) \, \omega(\alpha,v)
	\label{der5phif}
	\, .
\end{align}
As can be appreciated in the above examples, this line of attack quickly becomes intractable ---  in the next section we follow an alternative approach that simplifies the calculation of higher order derivatives of the geometric phase. 
\subsection{Derivatives of the geometric phase in terms of integrals}
\label{Dai}
Let $\omega$ be a time-independent $p$-form defined over a manifold.
Consider the integral
\begin{equation}
	\label{Italpha}
	I_t=\int_{V_t} \omega
	\, ,
\end{equation}
where the domain of integration $V_t$ is flowing (as $t$ varies) along the integral curves of a vector field $u$, $V_t=\phi_t^u(V_0)$,
with $\partial_t \phi_t^u (x)|_{t=0}=u(x)$. It can be shown that (see, \eg, \cite{Fra:04}),
\begin{equation}
	\label{ptIt}
	\partial_t I_t =\int_{V_t} L_u \omega
	\, ,
\end{equation}
where $L_u$ is the Lie derivative along $u$. We intend to use this formula to get an expression for the various time derivatives of the geometric phase --- a  toy example illustrating the use of~(\ref{ptIt}) appears in appendix \ref{Ate}.

Recall that
$\varphi_t$ is the usual Berry phase of the closed curve $c \circ h$ (with $\circ$ denoting concatenation), where $c_{t}$ denotes the curve $\rho_{\tau}$, $0 \leq \tau \leq t$,
and  $h_{t}$ is the geodesic that connects $\rho_0$ with $\rho_t$. By using the fact that the
symplectic form $\omega$ is proportional to the Berry curvature \cite{Ash.Sch:99}, we obtain (with the 
conventions for $\omega$ adopted above)
\begin{equation}
	\label{phiell}
	\varphi_t=-2\int_{V_t} \omega
	\, ,
\end{equation}
where $V_{t}$ is any surface with boundary $c \circ h$ --- we choose
as $V_t$  the surface swept out by the geodesics $h_\tau$, $0 \leq \tau \leq t$.

\begin{window}[0,r,%
	\includegraphics[width=.40\textwidth]%
	{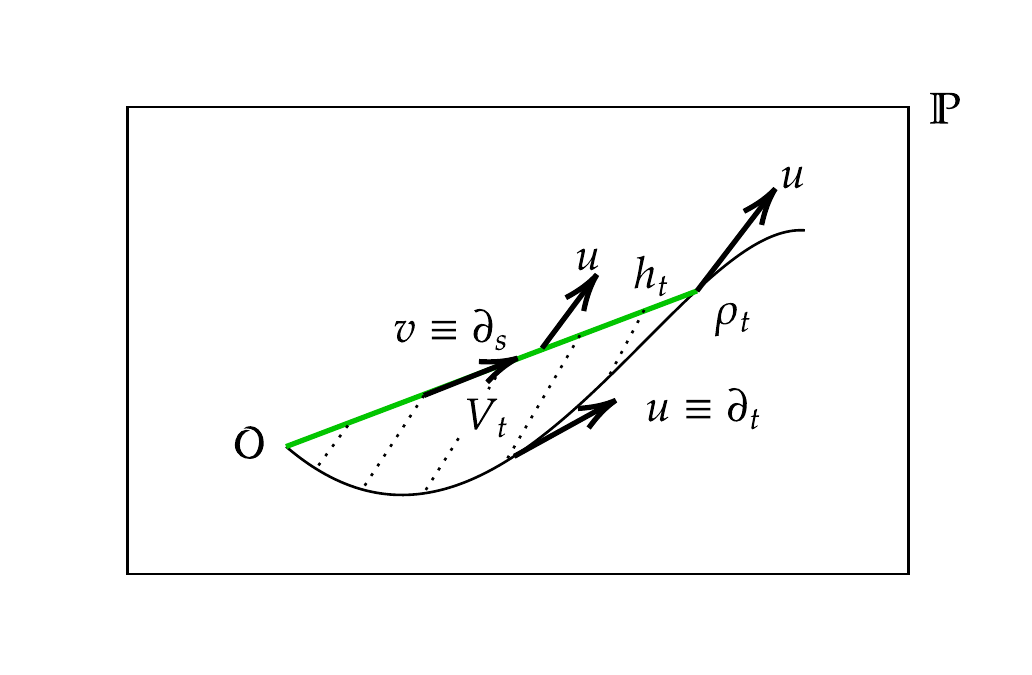},%
	{}]
	In the sketch on the right, the black curve denotes  $c_t$, while the green curve is $h_t$ --- we take the latter  parametrized by an affine parameter $s$, $0 \leq s \leq 1$, for all $t$. Assuming that the various geodesics
	$h_{t}$ for distinct values of $t$ only intersect at $\rho_{0}$,
	we can use the coordinates $(t,s)$ , $0 \leq t \leq T, 0 \leq s \leq 1$, to label the point corresponding to $s$ on $h_t$ ---
	the surface $V_t$, which is the hatched  area in the
	figure, corresponds to the range  $0 \leq \tau \leq t, 0 \leq s \leq 1$ and its boundary is 
	$\partial V_t=c_t-h_t$. Define the tangent vectors
	$u\equiv\partial_t$, $v \equiv \partial_s$. 
	Note that parametrizing $h_t$ by the affine parameter $s$ implies
	that all the points on $c_t$ have $s=1$, so that the vector field $u$ is tangent to
	$c_{t}$. Also, note that the surface $V_{t+dt}$ can be
	obtained by flowing the points of $V_{t}$ along the integral curves of $u$, as assumed in~(\ref{ptIt}).
\end{window}

Since $\omega$ is closed, by Cartan's formula, $L_u \omega = i_{u} d \omega + d (i_{u} \omega)=d (i_{u} \omega) $
(where $i_{u}$ denotes the contraction with $u$) holds, giving,
\begin{equation}
	\begin{split}
		\partial_t \varphi_t 
		=
		-2\int_{V_t} d(i_{u}\omega)
		=
		-2\int_{c_t} i_{u}\omega+2 \int_{h_t} i_{u}\omega 
		=
		2 \int_{h_t} i_{u}\omega,
	\end{split}
\end{equation}
where we used Stokes' theorem  for the second equality and
the fact that $i_{u}\omega=0$ at the curve $c_{t}$, since the corresponding line element is along
$u$  by construction.
Note that the integration in the above formula is only along the geodesic $h_t$ --- the curve $c_t$
affects the result via the vector field $u$, which depends on it.
By parametrizing $h_t$ with the coordinates $(\tau=t,s)$, $0 \leq s \leq 1$, we obtain
\begin{equation}
	\label{derellphi}
	\partial_t \varphi_t=2\int_{0}^{1} \td s \omega (u(s,t),v(s,t))
	\, ,
\end{equation}
and by computing the time derivative $k$ times,
\begin{equation}
	\partial_t^{k+1} \varphi_t
	=
	2\partial_t^{k} \int_0^1 \td s \, \omega(u(s,t),v(s,t))
	\, .
	\label{dervarphi2}
\end{equation}
As shown below, the integral on  the \rhs{}
of the above equation can be computed exactly.

Given a point $\rho_t$, consider the state $\ket{\psi_t}$ such that $\rho_t=\ket{\psi_t}\bra{\psi_t}$ and
$\braket{\psi_0}{\psi_t}=\cos L_t \geq 0$, where $L_{t}$ is the distance between $\rho_0$ and $\rho_t$.
Then the
geodesic $h_{t}$ can be parametrized by $s$ as follows
\begin{equation}
	\begin{split}
		h_{t}(s) = \ket{\psi_{ts}} \bra{\psi_{ts}}
	\end{split}
\end{equation}
where
\begin{equation} 
	\label{psiells}
	\ket{\psi_{t s}}=\cos (L_t s) \ket{\psi_0}+\sin(L_t s)\ket{\xi_t}
	\, ,
\end{equation}
with
\begin{equation}
	\label{phiell2}
	\ket{\xi_t}=\frac{1}{\sin L_t} \left(
	\ket{\psi_t}-\cos L_t \ket{\psi_0}
	\right)
\end{equation}
being the point of $h_t$ orthogonal to $\ket{\psi_0}$. Defining $\rho_{t s}=h_t(s)$ we may write
\begin{equation}
	\label{UaP}
	\rho_{t s}=U_{t s} \rho_0 U_{t s}^{-1}
	\, ,
\end{equation}
with $U_{ts}$ unitary.
Given $\rho_0$ and the point $\rho_{t s}$, (\ref{UaP}) does not determine $U_{t s}$ uniquely.
We fix this ambiguity by choosing $U_{t s}$, for fixed $t$, to be the one-parameter subgroup,
\begin{equation}
	\label{Uellsops}
	U_{t s}=e^{-i s L_{t} \xi_{t}}
	\, ,
	\qquad
	\chi_t \equiv  i(\ket{\xi_t} \bra{\psi_0}-\ket{\psi_0} \bra{\xi_t})
	\, .
\end{equation}
It is easily seen that
\begin{equation}
	\label{Uellspsi0}
	U_{t s}\ket{\psi_0}=\ket{\psi_{t s}}
	\, ,
\end{equation}
with $\ket{\psi_{t s}}$ defined in~(\ref{psiells}). Note that $U_{ts}$ evolves $\ket{\psi_0}$ along a geodesic (for fixed $t$ and varying $s$).
From~(\ref{UaP}) we get
\begin{equation}
	u(t,s)=\dot{\rho}_{t s}=-i [\hat{X}_{t s},\rho_{t s}]
	\, ,
	\qquad
	v(t,s)=\rho'_{t s}=-i [\hat{Y}_{t},\rho_{t s}]
	\, ,
	\label{rhodotp}
\end{equation}
where a prime denotes the partial derivative $\partial_{s}$
and
\begin{equation}
	\label{XYdef}
	\hat{X}_{t s} \equiv i \, \dot{U}_{t s} U^{-1}_{t s}
	\, ,
	\qquad
	\hat{Y}_{t} \equiv i \, U'_{t s} U^{-1}_{t s} = L_{t} \chi_{t}
	\, .
\end{equation}
\label{}
Substitution in~(\ref{derellphi}) gives
\begin{align}
	\partial_t \varphi_t
	&=
	\frac{1}{2i} \int_0^1 \td s 
	\Tr(
	\rho_{t s} [[\hat{X}_{t s},\rho_{t s}],[\hat{Y}_{t s},\rho_{t s}]]
	)
	=
	\frac{1}{2i} \int_0^1 \td s 
	\Tr(
	\rho_{t s} [\hat{X}_{t s},\hat{Y}_{t s}])
	=
	\frac{1}{2i} \int_0^1 \td s 
	\Tr(
	\rho_0 [X_{t s},Y_{t}])
	\, ,
	\label{dervphi3}
\end{align}
where
\begin{equation}
	\label{XYdef2}
	X_{t s} \equiv i \, U_{t s}^{-1} \dot{U}_{t s}
	\, ,
	\qquad
	Y_{t} \equiv i \, U_{t s}^{-1} U'_{ts} = L_{t}\chi_{t}
	\, .
\end{equation}

We calculate now $\hat{X}_{t s}$ explicitly.
We start  with the following expression for $U_{t s}$,
\begin{equation}
	\label{Uclosed}
	U_{t s}= I +\sin L_{t}s \,  \chi_t +(1-\cos L_{t} s) \chi_{t}^{2}
	\, ,
\end{equation}
which is easily derived by noting that $(-i \chi_t)^3=i\chi_t$,
implying that the eigenvalues of $-i \chi_t$ are $0$, $\pm i$,
and that $U_{ts}$ can be written as $U_{t s}=e^{-i s L_{t} \chi_t}= I + b \chi_t+ c \chi_t^2$, with  $b$, $c$
determined by substitution of the above eigenvalues for $\chi_t$. A straightforward calculation gives a somewhat lengthy expression for $[X_{t s},Y_{t s}]$ (see~(\ref{XYcommcomp})
of appendix \ref{Amoi}) --- in projecting this result onto $\rho_0$, only terms proportional to $\rho_0$ contribute, so that
\begin{align}
	\Tr \left(
	\rho_0 \left[X_{t s},Y_{t s} \right]
	\right)
	&=
	-L_t b_t \sin 2 L_{t}s
	\, ,
	\label{dervarphif}
\end{align}
where we defined $b_t \equiv \braket{\xi_t}{\dot{\xi}_t}$.
Finally, from the last equality of~(\ref{dervphi3}), we get
\begin{align}
	\partial_t \varphi_t
	=
	\frac{i}{4} 
	b_t (1-\cos 2L_t)
	=
	\frac{i}{2}
	b_t  \sin^2 L_t
	\, .
	\label{varphiellder2}
\end{align}

We cast now (\ref{varphiellder2}) in terms of the symplectic structure. To this end
we find that
\begin{align}
	\left[ Y_t, \partial_t Y_t \right]
	=
	L_t \left[ \chi_t, \partial_t (L_t \chi_t) \right]
	=
	L_t^2 \left[ {\chi}_t,\dot \chi_t \right]
	=
	L_t^2 \left(
	\ket{\xi_t} \bra{\dot{\xi}_t} -\ket{\dot{\xi}_t} \bra{\xi_t}+2 b_t \rho_0 \right),
	\label{YpY}
\end{align}
which projects to $\Tr\left( \rho_0 \left[ Y_t,\partial_t Y_t \right] \right)=2 L_t^2 b_t$.
Solving this for $b_t$ and substituting in the \rhs{} of~(\ref{varphiellder2}) gives,
\begin{align}
	\partial_t \varphi_t
	&=
	\frac{i}{4} \frac{\sin^2 L_t}{L_t^2} \Tr \left( \rho_0 \left[ Y_t,\partial_t Y_t \right] \right)
	\nonumber
	\\
	&=
	-\frac{1}{2} \frac{\sin^2 L_t}{L_t^2} \omega_{\rho_0} \left( Y_t,\partial_t Y_t \right)
	\nonumber
	\\
	&=
	-\frac{1}{2} \omega_{\rho_0} \left( \tilde{Y}_t,\partial_t \tilde{Y}_t \right),
	\label{pvphiYY}
\end{align}
where we used that $Y_{t}$ can be regarded as an element of $T_{\rho_0} \mathbb{P}$ (as it can be verified by noting
that its normal part is zero) to obtain the second line, and defined,
\begin{equation}
	\tilde{Y}_t \equiv \frac{\sin L_t}{L_t} Y_t = \sin L_t \chi_t
	\, .
\end{equation}
Higher order derivatives of the geometric phase are given by
\begin{equation}
	\partial_t^{k+1} \varphi_t
	= 
	-\frac{1}{2} \sum_{r=0}^k {k \choose r} \, 
	\omega_{\rho_0} \! 
	\left(
	\partial_t^r \tilde{Y}_t,\partial_t^{k-r+1} \tilde{Y}_t
	\right)
	\label{derphik1}
	\, .
\end{equation}
We conclude this section by noting that $Y_{t}$, regarded as an element of $T_{\rho_0} \mathbb{P}$,
has a precise geometrical interpretation;
a short calculation reveals that
the geodesic exponential map of $-J Y_{t}$ is $\rho_t$ by construction. Indeed, as noticed previously,
for fixed $t$ and varying $0 \leq s \leq 1$, $\rho_{ts}$ gives the geodesic from $\rho_0$ to $\rho_t$, and its tangent
vector at $s=0$ is $-J Y_{t}$, so $-J Y_{t}$ is the inverse of the exponential map of the curve $\rho_t$.
If we assume it has expansion of the form,
\begin{equation}
	\begin{split}
		- J Y_{t} = \sum_{n=1}^{\infty}  \tilde v^{(n-1)}\frac{t^{n}}{n!},
	\end{split}
\end{equation}
where $\tilde v^{(n)}$ is tangent vector in $T_{\rho_0} \mathbb{P}$, and
note that $\omega_{\rho_0}(\partial_t^r {Y}_t,\partial_t^{k} {Y}_t)=\omega_{\rho_0}(\partial_t^r J {Y}_t,\partial_t^{k} J {Y}_t)$, we obtain,
\begin{align}
		\partial_{t} \varphi_{0}
		&=
		0
		\\
		 \partial_{t}^{2}\varphi_{0}
		 &=
		 0
		 \\
		\partial_{t}^{3}\varphi_{0}
		&=
		 \omega( \tilde v^{(1)},\tilde v^{(0)})
		 \\
		\partial_{t}^{4}\varphi_{0}&=
		 2\omega(\tilde v^{(2)},\tilde v^{(0)}),
		 \\
		\partial_{t}^{5}\varphi_{0}
		&=
		3\omega(\tilde v^{(3)},\tilde v^{(0)})
		+\omega(\tilde v^{(2)},\tilde v^{(1)})
		- 2 g(\tilde v^{(0)},\tilde v^{(0)})\omega(\tilde v^{(1)},\tilde v^{(0)})
		\\
	 \partial_{t}^{6}\varphi_{0}
	 &=
		4 \omega(\tilde v^{(4)},\tilde v^{(0)}) +
		5 \omega( \tilde v^{(3)},\tilde v^{(1)}) -
		\frac{40}{3} g(\tilde v^{(0)},\tilde v^{(0)})\omega(  \tilde v^{(2)},\tilde v^{(0)}) -
		20 g(\tilde v^{(1)}, \tilde v^{(0)}) \omega( \tilde v^{(1)},\tilde v^{(0)})
		\, .
	\label{eq:GeoFirstFive}
\end{align}
The relation between $\tilde{v}^{(n)}$ and the covariant derivatives $(\nabla_v^{(g)})^n v$ is as follows
\begin{align}
		v 
		&= \tilde v^{(0)}
		\\
		\nabla_{v}^{(g)} v 
		&= 
		\tilde v^{(1)}
		\\
		(\nabla_{v}^{(g)})^{2} v 
		&=
		 \tilde v^{(2)}
		 \\
		(\nabla_{v}^{(g)})^{3} v 
		&=
		 \tilde v^{(3)}-g ( \tilde v^{(1)}, \tilde v^{(0)}) \tilde v^{(0)}
		-3 \omega(\tilde v^{(1)}, \tilde v^{(0)}) J \tilde v^{(0)}+ g(\tilde v^{(0)},\tilde v^{(0)}) \tilde v^{(1)}
		\, ,
\end{align}
and so on for higher values of $n$.

Finally, note that the tangent vectors $\tilde v^{(n)}$, $n\geq 1$, are trivially zero for geodesics,
making it clear  that all derivatives of the geometric phase (and hence, the phase itself) are zero in this case.


\section{The Brachistophase}
\label{Brachi}
We study Schr\"odinger curves in $\mathbb{P}$ that accumulate the maximum possible geometric phase for a given evolution time $\tau$.
\subsection{The hamiltonian of maximal acceleration}
\label{TurboH}
As a warmup, we consider the following problem: find the time-independent hamiltonian $H$ that, when used in Schr\"odinger's equation,  maximizes the initial acceleration of a given state $\rho_0$. From~(\ref{alphasqexpl}) we conclude that we need to maximize 
\begin{equation}
\label{fHdef}
f_{\rho_0}(H)=h_4-4h_3 h_1 -h_2^2+8 h_2 h_1^2 -4h_1^4
\, ,
\end{equation}
with $h_m \equiv \Tr(\rho_0 H^m)$ and where $f_{\rho_0}$ is viewed as a function from $\mathfrak{u}(n+1)$, where $H$ lives,  to the nonnegative reals. Since $f_{\rho_0}(\lambda H)=\lambda^4 f_{\rho_0}(H)$, we need to fix the norm of $H$ to, \eg, unity, to get a well-posed problem. Also, any component of $H$ along the unit matrix does not contribute to the dynamics of $\rho$, so the solution to our problem should have zero such component --- we arrive then at the following two constraints
\begin{align}
\Tr H
&=
0
\, ,
\label{TrH}
\\
\frac{1}{2} \Tr H^2
&=
1
\, .
\label{TrH2}
\end{align}
Both $f_{\rho_0}$ and the constraints are invariant under the transformation 
\begin{equation}
\label{UactrhoH}
\rho \rightarrow \rho'=U \rho \, U^{-1}
\, ,
\quad
H \rightarrow H'=UHU^{-1}
\, ,
\end{equation}
with $U \in U(n+1)$. Since the above action of $U(n+1)$ on $\mathbb{P}$ is transitive, we can solve the problem for any conveniently chosen state $\rho_0$, and then transform the solution as above, to solve it for any other state $\rho$. We choose then as $\rho_0$ the coherent state along $z$, and write accordingly the hamiltonian in the form
\begin{equation}
\label{Hdecomp}
H=\left(
\begin{array}{cc} b & v^\dagger
\\
v & B
\end{array}
\right)
\, ,
\end{equation}
where $b \in \mathbb{R}$ , $v \in \mathbb{C}^n$, and $B \in \mathfrak{u}(n)$. The stability subgroup $G_{\rho_0}$ of the above $\rho_0$ consists of matrices of the form
\begin{equation}
\label{Grho0}
V=\left(
\begin{array}{cc}
e^{i \alpha} & 0
\\
0 & W
\end{array}
\right)
\, ,
\end{equation}
with $W \in U(n)$. Under a transformation by such a matrix, $\rho_0$ remains invariant while $H$ transforms to $H'$ with 
\begin{equation}
\label{HtransG}
b'=b
\, ,
\quad
v'=e^{-i\alpha} Wv
\, ,
\quad
B'=WBW^{-1}
\, ,
\end{equation}
and the solution space for $H$, for a given $\rho_0$, is the entire orbit $G_{\rho_0} \triangleright H_0$ of a particular solution $H_0$ under $G_{\rho_0}$, together with the orbit of $-H_0$, since the latter hamiltonian clearly produces the same (modulus of) acceleration.

Using the above  form for $H$, we find 
\begin{align}
h_1
&=
b
\label{h1res}
\\
h_2
&=
b^2+\beta^2
\label{h2res}
\\
h_3
&=
b^3+2 b \beta^2+ v^\dagger B v
\label{h3res}
\\
h_4
&=
b^4+3b^2 \beta^2+2 b v^\dagger B v +v^\dagger B^2 v+\beta^4
\label{h4res}
\, ,
\end{align}
where  $\beta^2 \equiv v^\dagger v$, with $\beta \geq 0$, which leads to
\begin{equation}
\label{fH2}
f_{\rho_0}(H)=v^\dagger (B-b I)^2 v = |(B-b I)v|^2 \equiv |\tilde{B}v|^2
\, ,
\end{equation}
while the constraints, expressed in terms of $\tilde{B}$, assume the form
\begin{align}
(n+1)b+\Tr \tilde{B}
&=
0
\, ,
\label{TrHp}
\\
\frac{1}{2} b^2 +\frac{1}{2} \Tr \left( (\tilde{B}+b I)^2 \right) +\beta^2
&=
1
\label{TrH2p}
\, .
\end{align}
We may use a $G_{\rho_0}$ transformation (as in~(\ref{HtransG})) to bring $B$  (and, hence, $\tilde{B}$) in a diagonal form, $\tilde{B}=\text{diag}(\lambda_1,\ldots,\lambda_n)$.
To maximize the modulus of $\tilde{B}v$, we need to align $v$ along the eigenvector corresponding to the maximal (in the absolute sense) $\tilde{B}$-eigenvalue $\lambda$ and then make $|\lambda|\beta$ (which gives the modulus of the resulting vector) the maximum possible. We may assume, without loss of generality,  that the maximal, in absolute sense, eigenvalue of $\tilde{B}$ is $\lambda_1$.  The parameter space of the maximization problem is $\mathbb{R}$, where $b$ ranges, times $\mathfrak{u}(n)$, where $B$ lives, times $\mathbb{C}^n$, where $v$ lives, modulo the constraints. Using  Lagrange multipliers ($\mu_1$, $\mu_2$) to incorporate the constraints, we get
\begin{equation}
	\label{facc}
	f_{\rho_0}(H)=\lambda_1 \beta+\mu_{1}\left(
	\textup{Tr}\left(
	\tilde{B}+bI
	\right)^2
	-2+b^2+\beta^2
	\right)+\mu_2\left(
	\textup{Tr}\tilde{B}+(n+1)b
	\right)=f(\lambda_1,\dots,\lambda_n,\beta,b,\mu_1,\mu_2)
	\, .
\end{equation}
Now we compute the derivatives of $f$ with respect to every variable (fixing the other variables) and we get
\begin{align}
	\label{1c}
	\frac{\partial f}{\partial\lambda_r}&=2\mu_1(\lambda_r+b)+\mu_2=0,\qquad r\neq 1,\\
	\label{2c}
	\frac{\partial f}{\partial \lambda_1}&=\beta+2\mu_1(\lambda_1+b)+\mu_2=0,\\
	\label{3c}
	\frac{\partial f}{\partial b}&=2\mu_1 \text{Tr}\tilde{B}+2\mu_1 b+2\mu_1 n b+(n+1)\mu_2=0,\\
	\label{4c}
	\frac{\partial f}{\partial \beta}&=\lambda_1+4\mu_1\beta=0,\\
	\label{5c}
	\frac{\partial f}{\partial \mu_1}&=\text{Tr}\left(
	\left(\tilde{B}+bI\right)^2
	\right)-2+b^2+2\beta^2=0,\\
	\label{6c}
	\frac{\partial f}{\partial\mu_2}&=\text{Tr}\tilde{B}+(n+1)b=0,
\end{align}
where we consider that, for fixed $v$ and $b$,  $\tilde{B}$ has to lie on the $\Tr \tilde{B}=-(n+1)b$ hyperplane in $\mathfrak{u}(n+1)$,  satisfy additionally
\begin{equation}
	\label{TrH2cond}
	\Tr \left( (\tilde{B}+b I)^2 \right) =(\lambda_1+b)^2+\ldots+(\lambda_n+b)^2=2-b^2-2\beta^2
	\, ,
\end{equation}
and also have its first eigenvector along $v$. Substituting $\text{Tr}\tilde{B}$ from (\ref{6c}) in (\ref{3c}) we find  that $\mu_2=0$. Now substituting $\mu_2=0$ in (\ref{1c}) we obtain
\begin{equation}
	2\mu_1(\lambda_1+b)=0,\qquad r\neq 1.
\end{equation}
If $\mu_1=0$, by (\ref{2c}) we get that $\beta=0$, and then $f=0$ (its the minimum acceleration), hence we assume that $\mu_1\neq 0$ and $\lambda_r=-b$ for $r=2,\dots,n$. Now from (\ref{6c}) we have $\lambda_1-(n-1)b+(n+1)b=\lambda_1+2b=0$, thus $\lambda_1=-2b$. Substituting in (\ref{2c}) we get that $\beta=-2\mu_1(\lambda_1+b)=-2\mu_1(-2b)=2\mu_1b$. Then, from (\ref{4c}), we have
\begin{equation}
	\lambda_1+4\mu_1\beta=-2b+4\mu_1(2\mu_1b)=0\quad \rightarrow 2b(-1+4\mu_1^2)=0.
\end{equation}
If $b=0$, $H=0$, then we assume that $b\neq 0$ and $-1+4\mu_1^2=0$, \ie{}  $\mu_1=\pm\frac{1}{2}$, and, accordingly, $\beta=\pm b$. Finally, from (\ref{5c}), we get
\begin{equation}
	b^2-2+b^2+2b^2=0\quad \rightarrow\quad b=\frac{1}{\sqrt{2}}.
\end{equation}
Thus, the general form of $H$ that maximizes the initial acceleration of $\rho_0$ is given by

\begin{equation}
	\label{Hacc}
	H=\frac{1}{\sqrt{2}}
	\left(
	\begin{array}{ccccc}
		\pm 1 & 1 & 0 & \ldots & 0
		\\
		1 & \mp 1 & 0 & \ldots & 0
		\\
		0 & 0 & 0 & \ldots & 0
		\\
		\vdots  & \vdots & \vdots &  \ddots & \vdots
		\\
		0 & 0 & 0 & \ldots & 0
	\end{array}
	\right)
	\, ,
\end{equation}

\begin{myexample}{Maximum acceleration for a spin-1/2 state}{maxs12:Ex}
	For $s=\frac{1}{2}$ we have the maximum value of $f$ given by $\alpha^{(\frac{1}{2})}_{\text{max}}=1$, for  $H_{\frac{1}{2}}=\frac{1}{\sqrt{2}}(\pm \sigma_z+\sigma_x)$, \ie, in physical terms, to  a magnetic field at an angle of 45 (or 135) degrees \wrt{} the $z$-axis. The stability subgroup action rotates the direction of the magnetic field around the $z$-axis.
\end{myexample}
\subsection{The brachistophase hamiltonian}
\label{Hbra:subsec}
\subsubsection{Statement of the brachistophase problem}
\label{Sotbp}
We now consider the following problem: given an initial state $\rho_0$, find a time-independent hamiltonian $H$, such that, after a fixed time $\tau$ of evolution generated by $H$,  the geometric phase $\varphi_{\ell(\tau)}$ accumulated by the state is maximal. Since the phase is only defined modulo $2\pi$, we assume $t$ is sufficiently small for the phase to be less than $\pi$ at all times. Define the function $\Phi_{H,\, \rho_0}(t)$ to be the geometric phase accumulated by the initial state $\rho_0$ when it is evolved for time $t$ by the hamiltonian $H$. 
A Taylor expansion of $\Phi_{H,\, \rho_0}$ around $t=0$ gives
\begin{equation}
	\Phi_{H,\, \rho_0}(\tau)
	=
	\sum_{k=3}^\infty \frac{\tau^k}{k!} \partial_t^k \varphi_t|_{t=0} 
	\label{TaylorPhi}
	\, ,
\end{equation}
where we used the fact that $\varphi_t$ and its first two derivatives at $t=0$ vanish, while its higher-order derivatives can be computed with the help of~(\ref{derphik1}).
\subsubsection{Truncation up to $k=3$}
\label{Tutk3}
\begin{figure}[h]
	\centering
	\includegraphics[scale=0.50]{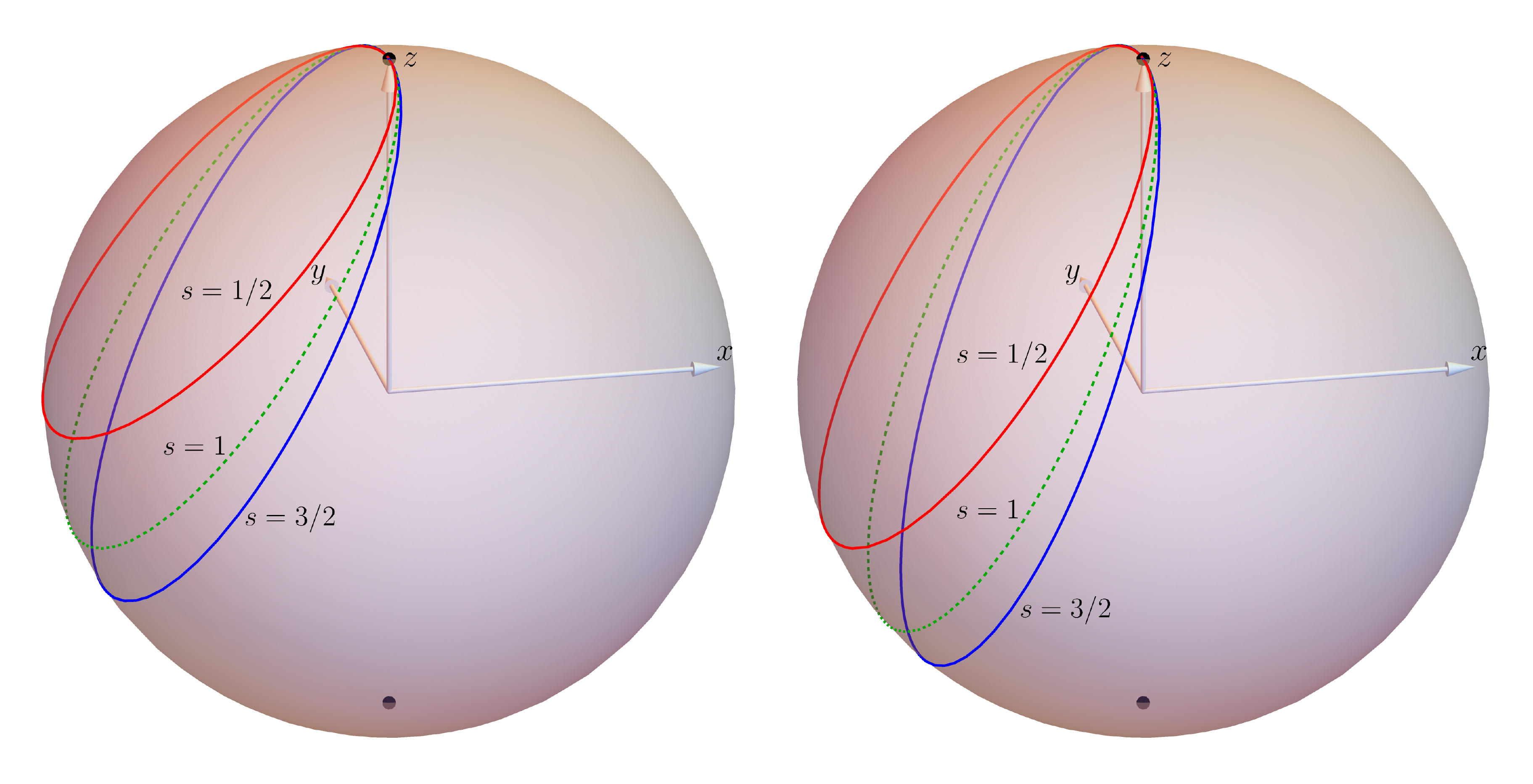}
	\caption{Falling stars for the cases $s=1/2,\ 1,\ 3/2$ for maximum acceleration (left) and for the brachistophase problem (right).}
	\label{GHZtetraSol:Fig}
\end{figure}

Truncating the expansion to only include the $k=3$ term, we need to maximize $\partial_t^3 \varphi_{t}|_{t=0}$. Note that~(\ref{varphiddd0H}) gives directly the third time derivative of the phase, so,  proceeding as before, we find (to order $\tau^3$)
\begin{align}
	\Phi_{H,\, \rho_0}(\tau)
	&=
	\frac{\tau^3}{6} \partial_t^3  \varphi_{t}|_{t=0} 
	\nonumber
	\\
	&=
	\frac{\tau^3}{6} (h_3-3h_2 h_1+2h_1^3)
	\nonumber
	\\
	&=
	\frac{\tau^3}{6}  v^\dagger \tilde{B} v 
	\nonumber
	\\
	&=
	\frac{\tau^3}{6} \beta^2 \lambda_1
	\label{Phitrunc3}
	\, ,
\end{align}
where, in the last step, $v$ has been assumed aligned as before. Using the Lagrange multipliers method as before we have

\begin{equation}
	\label{gbra}
	g_{\rho_0}(H)=\frac{\beta^2}{6}\lambda_1+\mu_1\left(
	\text{Tr}\left(
	\tilde{B}+bI
	\right)^2-2+b^2+2\beta^2
	\right)+\mu_2\left(
	\text{Tr}\tilde{B}+(n+1)b
	\right)=g(\lambda_1,\dots,\lambda_n,\beta,b,\mu_1,\mu_2).
\end{equation}
Calculating the partial derivatives of $g$ with respect to every variable we note that (\ref{1c}), (\ref{3c}), (\ref{5c}) and (\ref{6c}) are still valid, thus $\mu_2=0$ still holds, and additionally we have
\begin{align}
	\label{9c}
	\frac{\partial g}{\partial \lambda_1}&=\frac{\beta^2}{6}+2\mu_1(\lambda_1+b)+\mu_2=0,\\
	\label{10c}
	\frac{\partial g}{\partial \beta}&=\frac{1}{3}\beta\lambda_1+4\mu_1\beta=0,
\end{align}
from (\ref{1c}) and (\ref{9c}) we conclude that $\lambda_r=-b$, for $r=2,\dots,n$, and from (\ref{6c}) we have $\lambda_1=-2b$. Now using (\ref{9c}) we obtain
\begin{equation}
	\frac{\beta^2}{6}=-2\mu_1(\lambda_1+b)=2\mu_1b\quad\rightarrow\quad \beta=\sqrt{12\mu_1b}.
\end{equation} 
Substituting in (\ref{10c}) we get
\begin{equation}
	\frac{1}{3}\sqrt{12\mu_1b}(-2b)+4\mu_1\sqrt{12\mu_1b}=0\quad\rightarrow \quad \sqrt{12\mu_1b}\left(-\frac{2}{3}b+4\mu_1\right)=0,
\end{equation}
and, since we know that $\mu_1$ and $b$ are not 0, we have $\mu_1=\frac{1}{6}b$, and finally, substituting in (\ref{5c}) leads to
\begin{equation}
	b^2-2+b^2+4b^2=0\quad\rightarrow\quad b=\frac{1}{\sqrt{3}}.
\end{equation}
Therefore, the general form of $H$, solution to the brachistophase problem, is given by

\begin{equation}
	\label{Hbra}
	H=\frac{1}{\sqrt{3}}
	\left(
	\begin{array}{ccccc}
		\mp 1 & \sqrt{2} & 0 & \ldots & 0
		\\
		\sqrt{2} & \pm 1 & 0 & \ldots & 0
		\\
		0 & 0 & 0 & \ldots & 0
		\\
		\vdots  & \vdots & \vdots &  \ddots & \vdots
		\\
		0 & 0 & 0 & \ldots & 0
	\end{array}
	\right)
	\, ,
\end{equation}
\begin{myexample}{Spin-1/2 brachistophase}{brachis12:Ex}
	The corresponding hamiltonian for $s=1/2$ is
	
	\begin{equation}
		\label{Hs12phi}
		H^{(1)}_{\text{max}}=
		\frac{1}{\sqrt{3}} 
		\left(
		\begin{array}{cc}
			-1 & \sqrt{2}
			\\
			\sqrt{2} & 1
		\end{array}
		\right)
		\, ,
	\end{equation}
	where we use the negative root $b=-1/\sqrt{3}$, because in this way $\Phi_{H,\, \rho_0}$ is positive (actually there is no problem with the positive root due to we are interested in the absolute value of the geometric phase, but it is convenient to use positive quantities). The evolution under this hamiltonian describes a magnetic field with its axis at $\theta^{(\frac{1}{2})}=\arctan \sqrt{2} \approx 55^\circ$ \wrt{} the $z$-axis --- the trajectory of the Majorana star is a circle with its plane perpendicular to the field, see Fig.~\ref{GHZtetraSol:Fig} on p.{}~\pageref{GHZtetraSol:Fig}
	(more details about the Majorana representation of spin-$s$ states as 2$s$ points on the sphere may be consulted in~\cite{Maj:32,Chr.Guz.Ser:18}).
\end{myexample}
\begin{myexample}{Spin-$s$ brachistophase}{brachiss:Ex}
	Considering the optimal hamiltonian given by (\ref{Hbra}), the corresponding evolution operator is 
	\begin{equation}
		\label{Uts}
		U(t)=e^{-i t H}=
		\left(
		\begin{array}{ccccc}
			\cos t+\frac{i}{\sqrt{3}} \sin t & -i \sqrt{\frac{2}{3}} \sin t & 0 & \ldots & 0
			\\
			-i \sqrt{\frac{2}{3}} \sin t & \cos t-\frac{i}{\sqrt{3}} \sin t & 0 & \ldots & 0
			\\
			0 & 0 & 1 & \ldots & 0
			\\
			\vdots  & \vdots & \vdots &  \ddots & \vdots
			\\
			0 & 0 & 0 & \ldots & 1
		\end{array}
		\right)
		\, ,
	\end{equation}
	the first column of which gives the time-evolved state $\ket{\psi_t} \sim (1,\sqrt{2}/(-1+i \sqrt{3} \cot t),0,\ldots,0)^T$. The corresponding Majorana polynomial is $P_{\ket{\psi_t}}=\zeta^{2s}-\sqrt{2s} \sqrt{2}/(-1+i \sqrt{3} \cot t) \zeta^{2s-1}$, with a single nonzero root at $\zeta_t=2\sqrt{s}/(-1+i \sqrt{3}\cot t)$. Note that $\zeta_t-\sqrt{s}$ has modulus equal to $\sqrt{s}$ (independent of $t$) so $\zeta_t$ traces a circle in the complex plane with center at $\sqrt{s}$ and radius $\sqrt{s}$. It follows that the Majorana constellation of $\ket{\psi_t}$ consists of $2s-1$ stars at the north pole and a single ``falling star'' which, being the stereographic projection of $\zeta_t$, also traces a circle on the sphere --- the perpendicular to the plane of that circle makes an angle $\theta^{(s)}=\arctan (2\sqrt{s})$ \wrt{} the $z$-axis. 
\end{myexample}

Analogously it is possible to find that the evolution of quantum states under the hamiltonian that maximizes acceleration also consists of $2s-1$ stars at the north pole and a falling star which also traces a circle, but in this case the perpendicular to the plane of that circle makes an angle greater that the corresponding with the brachistophase problem.

\subsubsection{How good is the third order approximation?}
\label{Hgtoa}

The solution for $H$ when higher order terms are included in the Taylor expansion of the geometric phase will depend on the time $\tau$. Note that for Schr\"odinger curves the even-order derivatives at $t=0$ vanish, so the next simplest example involves the fifth order term. Working with the third order approximation implies that the $H$ found to solve the brachistophase problem is a valid solution for small values of time --- the question now is when is $\tau$ small?  Truncating the expansion up to the next order gives
\begin{equation}
	\Phi_{H,\rho_0}(\tau)=\frac{\tau^3}{3!}\partial_t^3\varphi_{\ell(t)}\bigg|_{t=0}+\frac{1}{5!}\partial_t^5\varphi_{\ell(t)} \bigg|_{t=0}+\mathcal{O}\left(\tau^7\right) 
	\, .
\end{equation}
The third order approximation becomes inadequate when the term corresponding to the fifth derivative becomes comparable to that of the third derivative. We compute the time $\tau_0$ for the two terms to become equal,
\begin{equation}
	\alpha \tau_0^3=\beta\tau_0^5,
\end{equation}
where $\alpha=\frac{1}{3!}\partial_t^3\varphi_{t}$ and $\beta=\frac{1}{5!}\partial_t^5\varphi_{t}$. We get 
$\tau_0=\sqrt{\frac{\alpha}{\beta}}$,
which, for the optimal $H$, is equal to $\tau_0=1.732$ --- the above approximation is then good for $\tau\ll \tau_0$.

\begin{figure}[h]
	\centering
	\includegraphics[scale=1]{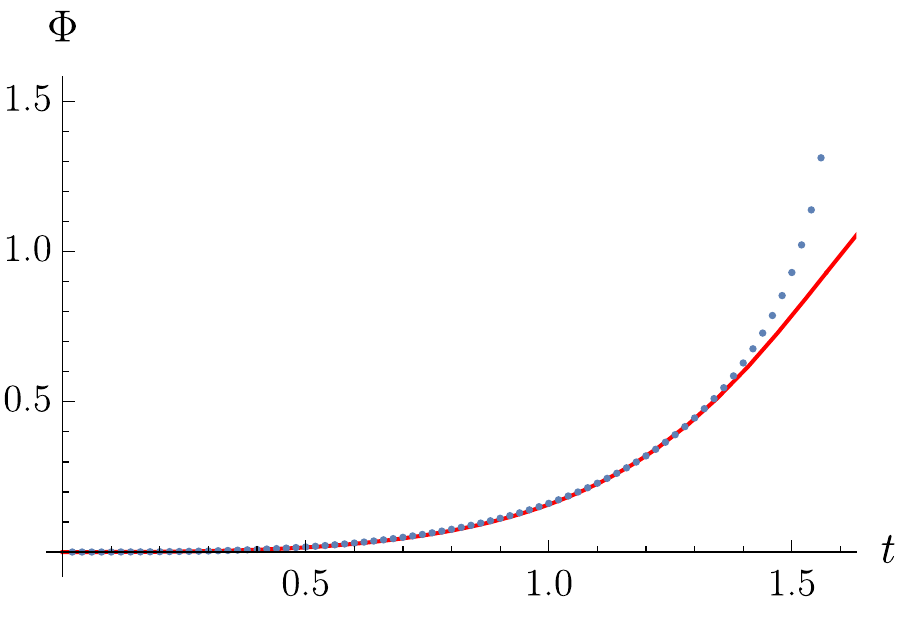}
	\caption{Geometric phase corresponding to the numerically determined optimal hamiltonian  (dotted blue  curve) and the third order approximation (continuous red curve). 
	}
	\label{truncvsnum:Fig}
\end{figure}

Maximizing the exact expression of the geometric phase through a set of random hamiltonians generated numerically, we see that for $t<1.5$ the third order approximation is excellent,  while for $t>1.5$ the exact (numerical) geometric phase starts becoming significantly larger than the truncated result, see Figure~\ref{truncvsnum:Fig}.

\subsubsection{Evolution of GHZ and tetrahedral states}
\label{EGHZT}

Using (\ref{UactrhoH}) we find the optimal hamiltonian for the spin-3/2 GHZ and spin-2  tetrahedral states. We have $\ket{\psi_{\text{GHZ}}}=\frac{1}{\sqrt{2}}\left(1, 0, 0, -1\right)^T$ and $\ket{\psi_{\text{tetra}}}=\frac{1}{\sqrt{3}}\left(1, 0,0,\sqrt{2},0\right)^T$. From (\ref{Uclosed}) we find

\begin{equation}
	U_{\text{GHZ}}=\left(
	\begin{array}{cccc}
		\frac{1}{\sqrt{2}} & 0 & 0 & \frac{1}{\sqrt{2}} \\
		0 & 1 & 0 & 0 \\
		0 & 0 & 1 & 0 \\
		-\frac{1}{\sqrt{2}} & 0 & 0 & \frac{1}{\sqrt{2}} \\
	\end{array}
	\right),\qquad
		U_{\text{tetra}}=\left(
		\begin{array}{ccccc}
			\frac{1}{\sqrt{3}} & 0 & 0 & -\sqrt{\frac{2}{3}} & 0 \\
			0 & 1 & 0 & 0 & 0 \\
			0 & 0 & 1 & 0 & 0 \\
			\sqrt{\frac{2}{3}} & 0 & 0 & \frac{1}{\sqrt{3}} & 0 \\
			0 & 0 & 0 & 0 & 1 \\
		\end{array}
		\right),
\end{equation}
resulting in
\begin{equation}
	H'_{\text{GHZ}}
	=
	\left(
	\begin{array}{cccc}
		-\frac{1}{2 \sqrt{3}} & \frac{1}{\sqrt{3}} & 0 & \frac{1}{2 \sqrt{3}} \\
		\frac{1}{\sqrt{3}} & \frac{1}{\sqrt{3}} & 0 & -\frac{1}{\sqrt{3}} \\
		0 & 0 & 0 & 0 \\
		\frac{1}{2 \sqrt{3}} & -\frac{1}{\sqrt{3}} & 0 & -\frac{1}{2 \sqrt{3}} \\
	\end{array}
	\right)
	\, ,
	\qquad
	H'_{\text{tetra}}
	=
	\left(
	\begin{array}{ccccc}
		-\frac{1}{3 \sqrt{3}} & \frac{\sqrt{2}}{3} & 0 & -\frac{\sqrt{\frac{2}{3}}}{3} & 0 \\
		\frac{\sqrt{2}}{3} & \frac{1}{\sqrt{3}} & 0 & \frac{2}{3} & 0 \\
		0 & 0 & 0 & 0 & 0 \\
		-\frac{\sqrt{\frac{2}{3}}}{3} & \frac{2}{3} & 0 & -\frac{2}{3 \sqrt{3}} & 0 \\
		0 & 0 & 0 & 0 & 0 \\
	\end{array}
	\right)
	\, .
	\label{HGHZtetra}
\end{equation}
The evolution of the above quantum states is given by
\begin{align}
\ket{\psi_{\text{GHZ}}(t)}
&=
\left(
\frac{\cos (t)}{\sqrt{2}}
+
\frac{i \sin (t)}{\sqrt{6}}
\, ,
-i \sqrt{\frac{2}{3}} \sin (t)
\, ,
0
\, ,
-\frac{\cos (t)}{\sqrt{2}}-\frac{i \sin
	(t)}{\sqrt{6}}
	\right)^T
	\label{GHZsolt}
	\\
	\ket{\psi_{\text{tetra}}(t)}
	&=
	\left(
	\frac{\cos (t)}{\sqrt{3}}+\frac{1}{3} i \sin (t)
	\, ,
	-i \sqrt{\frac{2}{3}} \sin (t)
	\, ,
	0
	\, ,
	\frac{1}{3} \left(\sqrt{6} \cos (t)+i \sqrt{2} \sin(t)
	\right)
,
0
\right)^T
\, ,
\label{tetrasolt}
\end{align}
with plots appearing in figures 3 and 4, respectively.

\begin{figure}[h]
	\centering
	\includegraphics[scale=0.35]{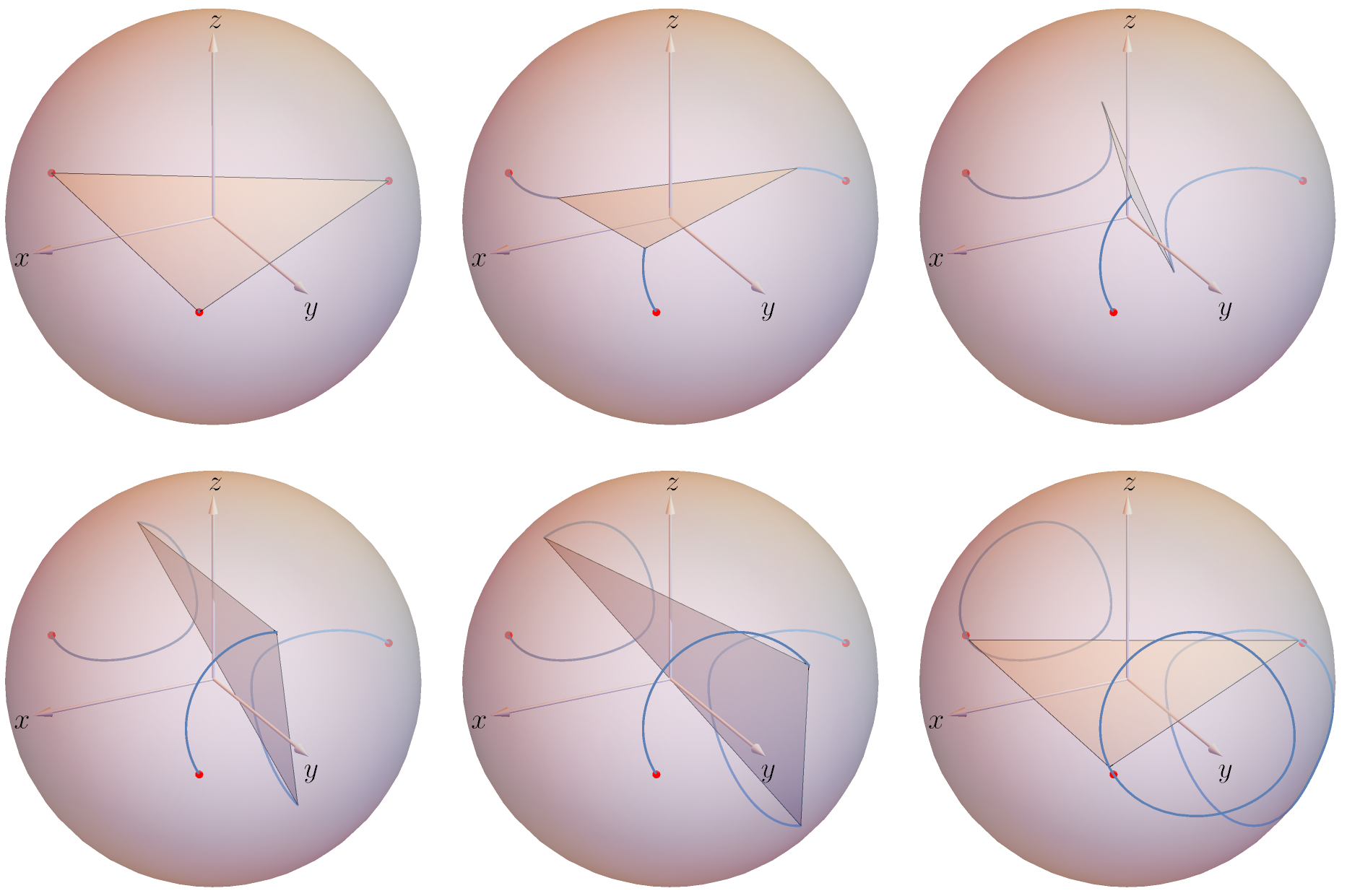}
	\caption{Optimal time evolution for a spin-3/2 GHZ state: shown is the Majorana constellation for $t=0,\ 0.5,\ 1,\ 1.5,\ 2,\ 3.2$ (left-to-right, top-to-bottom). The red points represent the original constellation, the curves in blue describe the trajectory of the stars during the evolution, while the triangles shown help visualize the constellation at the given time $t$.}
	\label{GHZSol:Fig}
\end{figure}

\begin{figure}[h]
	\centering
	\includegraphics[scale=0.35]{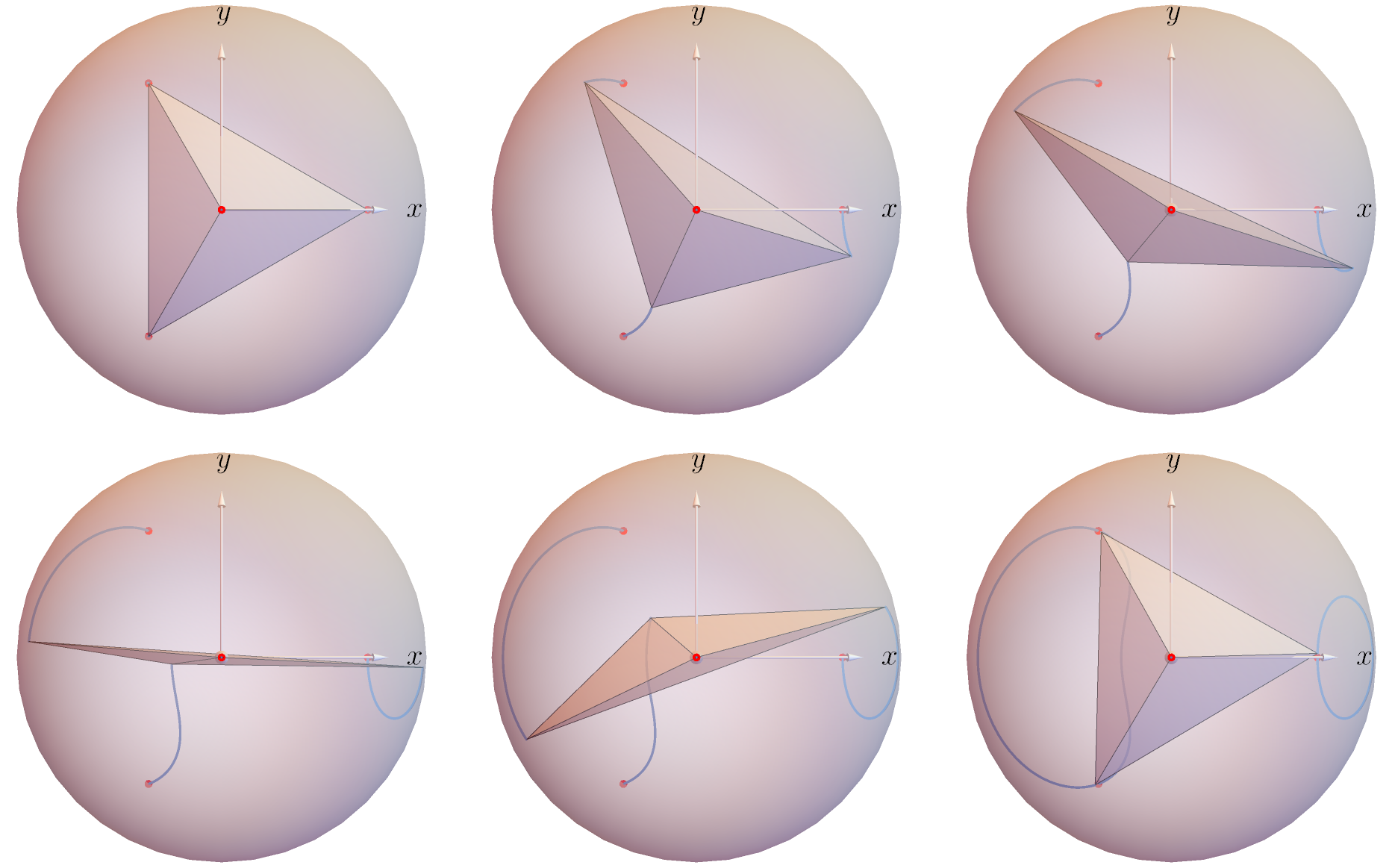}
	\caption{Optimal time evolution for the spin-2 tetrahedral state: shown is the Majorana constellation for $t=0,\ 0.5,\ 1,\ 1.5,\ 2,\ 3.2$ (left-to-right, top-to-bottom). The red points represent the original constellation, the curves in blue describe the trajectory of the stars during the evolution, while the  polyhedra shown help visualize the constellation at the given time $t$. Note that the star in the north pole (closest to the reader) is fixed, the one to the right  traces a circle, while the remaining two are permuted in the final frame.}
	\label{tetraSol:Fig}
\end{figure}
\section{Summary and concluding remarks}
\label{Conclusions}
We have studied the relation between geometric phase and covariant derivatives for a smooth curve in quantum state space. We found that the various derivatives of the geometric phase are proportional to the symplectic area of the parallelograms generated by various pairs of covariant derivatives, \eg, the first nonvanishing derivative of the phase (of the third order) is exactly equal to the symplectic area of the parallelogram generated by the velocity and the acceleration of the curve (see~(\ref{varphidddav})). When the curve in question corresponds to evolution generated by a time-independent hamiltonian, the time derivatives of the phase can be related to the expectation values of powers of the hamiltonian (see, \eg,~(\ref{varphiddd0H})). A general formula for the various time derivatives of the phase is given in~(\ref{derphik1}). It is worth emphasizing at this point that the geometric phase accumulated by a curve is not additive under curve concatenation, \eg, if a curve $c_1$, going from point $A$ to point $B$, is glued to a curve $c_2$, going from $B$ to $C$, the geometric phase for the resulting curve, going from $A$ to $C$,  is not the sum of the phases for the $c_i$. This implies that the phase derivatives mentioned above depend on the starting point, $t=0$, of the curve --- in our analysis said derivatives are calculated exactly at the starting point.

As an application of our geometric analysis, we discussed two maximization problems: given an initial state, find the (appropriately normalized) hamiltonian that maximizes i) the modulus of its initial acceleration and ii)  (the modulus of) the geometric phase accumulated after a fixed time $\tau$. 
Both problems were solved  with the initial state being a coherent state along $\hat{z}$ (see~(\ref{Hacc}), (\ref{Hbra})). For both problems, the solution for the maximizing hamiltonian is not unique --- given a particular solution, one obtains more solutions by acting on it with the stability subgroup of the initial state, while the time evolution of the state is independent of the particular optimal hamiltonian chosen.  Starting with a coherent state along $\hat{z}$, the time evolution generated by any optimal hamiltonian consists  of  a single star leaving the north pole and tracing out a circle, the characteristics of which are different for the two problems, and also depend on the spin of the state --- a few examples are depicted in Figure~\ref{GHZtetraSol:Fig}. The time evolution for other initial states is then easily obtained using the transitive action of the unitary group on the state space --- see~(\ref{HGHZtetra}) and Figures~\ref{GHZSol:Fig}, \ref{tetraSol:Fig}, for the brachistophase solution for the spin-3/2 GHZ  and the tetrahedral state. Note that the optimal hamiltonian  for the brachistophase problem depends, in general, on the time $\tau$. Our analytic solution is valid for appropriately defined small times, where the cubic term (in $\tau$) dominates. In this approximation the solution does not depend on $\tau$ --- including higher-order terms seems like a rather hard problem analytically and should probably be attempted with numerical methods. 

There are several open problems that we are currently pursuing as a follow up to the present paper's considerations. In particular, we would like to elucidate the physical significance of the modulus of the various covariant derivatives of a state, appropriately averaged over the driving hamiltonians. There is empirical evidence that  the functions on state space so defined can be used as measures of interesting physical properties, like entanglement (when the spin-$s$ state is considered as a symmetrized state of $2s$ spin-1/2 subsystems) --- this points to possible connections with the ``total variance'' concept in~\cite{Kly:07,Saw.Osz.Kus:12}.
\section*{Acknowledgements}
\label{Ack}
C.{} C.{} would like to acknowledge financial support from the DGAPA-UNAM project IN111920. E.{} S.{} E.{} acknowledges financial support from postdoctoral fellowships by DGAPA-UNAM and the IPD-STEMA Program of the University of Liège.

\appendix
\section{A miscellany of identities}
\label{Amoi}
We gather here several identities that we found useful in deriving the results of this paper. 

We compute
\begin{align}
U_{\ell s}^{-1}
&=
 I - \sin \tilde{s} \,  \chi_\ell -(1-\cos \tilde{s}) (\rho_0 +\sigma_\ell)
 \label{Uiexp1}
 \\
 \dot{U}_{\ell s}
 &=
 \dot{\tilde{s}} \cos \tilde{s} \, \chi_\ell+ \sin \tilde{s} \dot{\chi}_\ell 
 - \dot{\tilde{s}} \sin \tilde{s} (\rho_0 +\sigma_\ell) -(1-\cos \tilde{s}) \dot{\sigma}_\ell
 \label{Udexp1}
 \\
 U'_{\ell s}
 &=
 U_{\ell s} \, L_\ell \chi_\ell
 \, ,
 \label{Upexp1}
 \\
 U_{\ell s}^{-1} \dot{U}_{\ell s}
 &=
 \dot{\tilde{s}} \chi_\ell 
 +\sin \tilde{s} \, \dot{\chi}_\ell 
 -(1-\cos \tilde{s}) \dot{\sigma}_\ell
 +2(1-\cos \tilde{s}) \ket{\xi_\ell} \bra{\dot{\xi}_\ell}
 + \sin^2 \tilde{s} \, b_\ell \rho_0
 \nonumber
 \\
 & \qquad  
 -\sin \tilde{s}\,  (1-\cos\tilde{s}) b_\ell
 \tau_\ell
 +(1-\cos \tilde{s})^2 b_\ell \sigma_\ell
 \, ,
 \label{UiUdexp1}
 \\
 U_{\ell s}^{-1}U'_{\ell s}
 &=
 L_\ell \chi_\ell
 \, ,
 \label{UiUpexp1}
 \\
  \dot{U}_{\ell s}U_{\ell s}^{-1} 
 &=
\dot{\tilde{s}} \chi_\ell 
+\sin \tilde{s} \, \dot{\chi}_\ell 
 -(1-\cos \tilde{s}) \dot{\sigma}_\ell
 +2(1-\cos \tilde{s}) \ket{\dot{\xi}_\ell} \bra{\xi_\ell}
 - \sin^2 \tilde{s} \, b_\ell \rho_0
 \nonumber
 \\
 & \qquad  
 -\sin \tilde{s}\,  (1-\cos\tilde{s}) b_\ell \tau_\ell
    -(1-\cos \tilde{s})^2 b_\ell  \sigma_\ell
 \, ,
 \label{UdUiexp1}
 \\
 U'_{\ell s}U_{\ell s}^{-1}
 &=
 L_\ell \chi_\ell
 \, ,
 \label{UpUiexp1}
 \end{align}
  where $b_\ell \equiv \braket{\xi_\ell}{\dot{\xi}_\ell}=-\braket{\dot{\xi}_\ell}{\xi_\ell}$, and use was made of the identities
  \begin{alignat}{3}
  \chi_\ell \dot{\chi}_\ell
  &=
  -\ket{\xi_\ell} \bra{\dot{\xi}_\ell}-b_\ell \rho_0
  \, ,
  & \qquad 
  \dot{\chi}_\ell \chi_\ell
  &=
  -\ket{\dot{\xi}_\ell} \bra{\xi_\ell}+b_\ell \rho_0
  \label{id1}
  \\
  \chi_\ell \dot{\sigma}_\ell
  &=
  -b_\ell \ket{\psi_0} \bra{\xi_\ell} -\ket{\psi_0} \bra{\dot{\xi}_\ell}
  \, ,
  & \qquad 
  \dot{\sigma}_\ell \chi_\ell
  &=
 -b_\ell \ket{\xi_\ell} \bra{\psi_0}+\ket{\dot{\xi}_\ell} \bra{\psi_0}
  \, ,
  \label{id2}
  \\
  (\rho_0+\sigma_\ell) \dot{\chi}_\ell
  &=
  b_\ell \ket{\xi_\ell} \bra{\psi_0}- \ket{\psi_0} \bra{\dot{\xi}_\ell}
  \, ,
  & \qquad 
 \dot{\chi}_\ell  (\rho_0+\sigma_\ell) 
  &=
 b_\ell\ket{\psi_0} \bra{\xi_\ell}+\ket{\dot{\xi}_\ell} \bra{\psi_0}
  \, ,
  \label{id3}
  \\
  (\rho_0+\sigma_\ell) \dot{\sigma}_\ell
  &=
  b_\ell \sigma_\ell + \ket{\xi_\ell} \bra{\dot{\xi}_\ell}
  \, ,
  & \qquad 
\dot{\sigma}_\ell  (\rho_0+\sigma_\ell)
  &=
 -b_\ell  \sigma_\ell+\ket{\dot{\xi}_\ell} \bra{\xi_\ell}
  \, ,
   \label{id4}
  \\
  (\rho_0+\sigma_\ell)\chi_\ell
  &=
  \chi_\ell
  \, ,
   & \qquad 
   \chi_\ell (\rho_0+\sigma_\ell)
  &=
 \chi_\ell
  \, ,
  \label{id5}
  \\
  (\rho_0+\sigma_\ell)^2
  &=
  \rho_0+\sigma_\ell
  \, ,
  & \qquad 
  &
  \label{id6}
  \end{alignat}
  the last two of which express the fact  that $\rho_0+\sigma_\ell$ is a projection operator onto the plane spanned by $\ket{\psi_0}$, $\ket{\xi_\ell}$. We also compute some useful commutators,
  \begin{align}
  \left[ \dot{\chi}_\ell,\chi_\ell \right]
  &=
  \ket{\xi_\ell} \bra{\dot{\xi}_\ell}- \ket{\dot{\xi}_\ell} \bra{\xi_\ell} + 2 b_\ell  \rho_0
  \, ,
  \label{comm1}
  \\
  \left[ \tau_\ell,\chi_\ell \right]
  &=
  2(\rho_0-\sigma_\ell)
  \, ,
  \label{comm2}
  \\
  \left[  \ket{\xi_\ell} \bra{\dot{\xi}_\ell}, \chi_\ell \right]
  &=
  \ket{\psi_0} \bra{\dot{\xi}_\ell} -b_\ell \ket{\xi_\ell} \bra{\psi_0}
  \, ,
  \label{comm3}
  \\
  \left[  \ket{\dot{\xi}_\ell} \bra{\xi_\ell}, \chi_\ell \right]
  &=
  \ket{\dot{\xi}_\ell} \bra{\psi_0}+ b_\ell \ket{\psi_0} \bra{\xi_\ell}
  \, ,
  \label{comm3p}
  \\
  \left[ \rho_0, \chi_\ell \right]
  &=
  - \tau_\ell
  \, ,
  \label{comm4}
  \\
  \left[ \dot{\sigma}_\ell, \chi_\ell \right]
  &=
 \dot{\tau}_\ell- b_\ell \chi_\ell
  \, ,
  \label{comm5}
  \\
  \left[ \sigma_\ell, \chi_\ell \right]
  &=
  \tau_\ell
  \, ,
  \label{comm6}
  \end{align}
  so that
  \begin{align}
  -L_\ell^{-1} [X_{\ell s},Y_{\ell s}]
  &=
  [U_{\ell s}^{-1} \dot{U}_{\ell s}, \chi_{\ell}]
  \nonumber
  \\
  &=
  \sin \tilde{s} \left(
  \ket{\xi_\ell}\bra{\dot{\xi}_\ell} -\ket{\dot{\xi}_\ell} \bra{\xi_\ell} 
  +2b_\ell \rho_0
  \right)
  \nonumber
  \\
  &
  \qquad 
  {}-2 \sin \tilde{s} (1-\cos \tilde{s}) b_\ell (\rho_0-\sigma_\ell)
  \nonumber
  \\
  &
  \qquad
  {}+2 (1-\cos\tilde{s})
  \left(
   \ket{\psi_0} \bra{\dot{\xi}_\ell} -b_\ell \ket{\xi_\ell} \bra{\psi_0}
  \right)
  \nonumber
  \\
  &
 \qquad
  -\sin^2 \tilde{s} \, b_\ell \tau_\ell
  \nonumber
  \\
  &
  \qquad
  {}-(1-\cos \tilde{s}) 
  \left(
  \dot{\tau}_\ell-b_\ell \chi_\ell
  \right)
  \nonumber
  \\
  &
  \qquad
  {}+ (1-\cos \tilde{s})^2
 b_\ell  \tau_\ell
  \, ,
  \label{XYcommcomp}
  \end{align}
   and
  \begin{align}
  -L_\ell^{-1} [\hat{X}_{\ell s},\hat{Y}_{\ell s}]
  &=
  \sin \tilde{s} \left(
  \ket{\xi_\ell} \bra{\dot{\xi}_\ell} -\ket{\dot{\xi}_\ell} \bra{\xi_\ell}
  \right)
  + b_\ell \sin (2\tilde{s}) \rho_0
  +2 b_\ell \cos \tilde{s} (1-\cos \tilde{s}) \tau_\ell
  \nonumber
  \\
  & \quad
  -(1-\cos\tilde{s}) \dot{\tau}_\ell + b_\ell (1-\cos\tilde{s}) \chi_\ell
  + 2 (1-\cos\tilde{s}) \ket{\dot{\xi}}\bra{\psi_0} 
  \nonumber
  \\
  & \quad
  +2 b_\ell (1-\cos\tilde{s}) \ket{\psi_0} \bra{\xi_\ell} + 2 b_\ell \sin \tilde{s} ( 1-\cos\tilde{s}) \sigma_\ell
  \label{XYhatcomm}
  \, .
  \end{align} 
  We now compute the commutators
  \begin{alignat}{3}
  [\rho_0,\chi_\ell]
  &=
  -\tau_\ell
  \, ,
  & \qquad
  [\rho_0,\sigma_\ell]
&=
0
\, , 
  \label{commrho1}
\\
[\rho_0,\dot{\chi}_\ell]
&=
-\dot{\tau}_\ell
\, ,
& \qquad
[\rho_0,\dot{\sigma}_\ell]
&=
0
\, ,
\label{rhocomm2}
\\
[\rho_0,\tau_\ell]
&=
-\chi_\ell
\, ,
& \qquad
[\rho_0,\ket{\xi_\ell}\bra{\dot{\xi}_\ell}]
&=
0
\, ,
\label{rhocomm3}
\\
[\rho_0,\dot{\tau}_\ell]
&=
-\dot{\chi}_\ell
\, ,
& \qquad
[\rho_0,\ket{\dot{\xi}_\ell}\bra{\xi_\ell}]
&=
0
\, ,
\label{rhocomm4}
  \end{alignat}
  which allow the computation of the tangential part of $\hat{X}_{\ell s}$, $\hat{Y}_{\ell s}$ at $\rho_0$,
  \begin{align}
  -i \hat{X}_{\ell s}^{\parallel_0} &= [\rho_0,[\rho_0,\dot{U}_{\ell s} U^{-1}_{\ell s}]]
  = 
  \dot{\tilde{s}} \chi_\ell
  + \sin \tilde{s} \dot{\chi}_\ell -\sin \tilde{s} (1-\cos \tilde{s}) b_\ell \tau_\ell
  \, ,
  \\
   \hat{Y}_{\ell s}^{\parallel_0}
  &=
  \hat{Y}_{\ell s}
  \, ,
  \label{hatXYparallel}
  \end{align}
  with $X_{\ell s}^\parallel =\hat{X}_{\ell s}^\parallel$, $Y_{\ell s}^\parallel =\hat{Y}_{\ell s}^\parallel$. Note that the commutator of the tangential parts (at $\rho_0$)  reproduces exactly the normal part (at $\rho_0$) of the commutator in~(\ref{XYhatcomm}),
  \begin{equation}
  [\hat{X}_{\ell s}^{\parallel_0}, \hat{Y}_{\ell s}^{\parallel_0}]^{\perp_0} =  [\hat{X}_{\ell s}, \hat{Y}_{\ell s}]^{\perp_0}
  \, ,
  \end{equation}
  in accordance with our earlier discussion of even and odd parts, and the fact that $\hat{Y}_{\ell s}^{\perp_0}=0$.
  \begin{aside}
  Clarifying this last point, we have 
  \begin{align*}
  [X,Y]
  &=
  [X^\parallel+X^\perp,Y^\parallel + Y^\perp]
  \\
  &=
  \underbrace{[X^\parallel,Y^\parallel]+[X^\perp,Y^\perp]}_{[X,Y]^\perp}
  +\underbrace{[X^\parallel,Y^\perp]+[X^\perp,Y^\parallel]}_{[X,Y]^\parallel}
  \, ,
  \end{align*}
  where the sum of the first two terms gives $[X,Y]^\perp$ while the last two sum to $[X,Y]^\parallel$. Since $Y^\perp=0$, we get $[X,Y]^\perp=[X^\parallel,Y^\parallel]$.
  \end{aside}
  The above discussion relates to the tangent space at $\rho_0$. Referring to the first of~(\ref{dervphi3}) we would like to derive a similar statement at $\rho_{\ell s}$. To this end, note that if $\braket{a}{b}=0$, then 
  \begin{equation}
  \ket{a}\bra{b}-\ket{b}\bra{a}=\ket{\tilde{a}}\bra{\tilde{b}} - \ket{\tilde{b}}\bra{\tilde{a}}
  \quad \text{with} \quad 
  \ket{\tilde{a}}=\cos \alpha \ket{a}-\sin\alpha \ket{b}
  \, , \quad
  \ket{\tilde{b}}=\sin\alpha \ket{a} +\cos \alpha \ket{b}
  \, .
  \end{equation}
  Applying this to $\chi_\ell$ we get
  \begin{align}
  \chi_\ell
  &=
  \ket{\xi_\ell} \bra{\psi_0} -\ket{\psi_0}\bra{\xi_\ell}
  \nonumber
  \\
  &=
  \ket{\tilde{\xi}} \bra{\psi_{\ell s}}-\ket{\psi_{\ell s}} \bra{\tilde{\xi}}
  \, ,
  \label{chip}
  \end{align}
  with $\ket{\tilde{\xi}}=\cos \tilde{s} \ket{\xi_\ell}-\sin\tilde{s} \ket{\psi_0}$, 
  which shows that $\chi_\ell$ and, hence,  $\hat{Y}_{\ell s}$, are tangent to $\mathbb{P}$ at $\rho_{\ell s}$, for all $s$. Invoking the above argument we then get
  \begin{align}
  \Tr \left(
  \rho_{\ell s} \left[ \hat{X}_{\ell s}, \hat{Y}_{\ell s} \right]
  \right)
  &=
  \Tr \left(\rho_{\ell s} \left[ \hat{X}_{\ell s}, \hat{Y}_{\ell s} \right]^{\perp_{\ell s}} \right)
  \nonumber
  \\
  &=
   \Tr \left( \rho_{\ell s} \left[ \hat{X}_{\ell s}^{\parallel_{\ell s}}, \hat{Y}_{\ell s}^{\parallel_{\ell s}} \right]
   \right)
  \nonumber
  \\
  &=
  2i \omega_{\ell s} \left(\hat{X}_{\ell s}^{\parallel_{\ell s}}, \hat{Y}_{\ell s}^{\parallel_{\ell s}} \right)
  \label{TrXYh}
  \end{align}

\section {A toy example of equation (\ref{ptIt})}
\label{Ate}
In this appendix, we illustrate how to work with equation (\ref{ptIt}) with a very simple example: finding the area
of a circular arc in the plane.

Consider the circular arc in the sketch in the right, and take as $V_t$ the cross-hatched area in green, defined by the arc itself
\begin{window}[0,r,$\quad$%
	\includegraphics[width=.15\textwidth]%
	{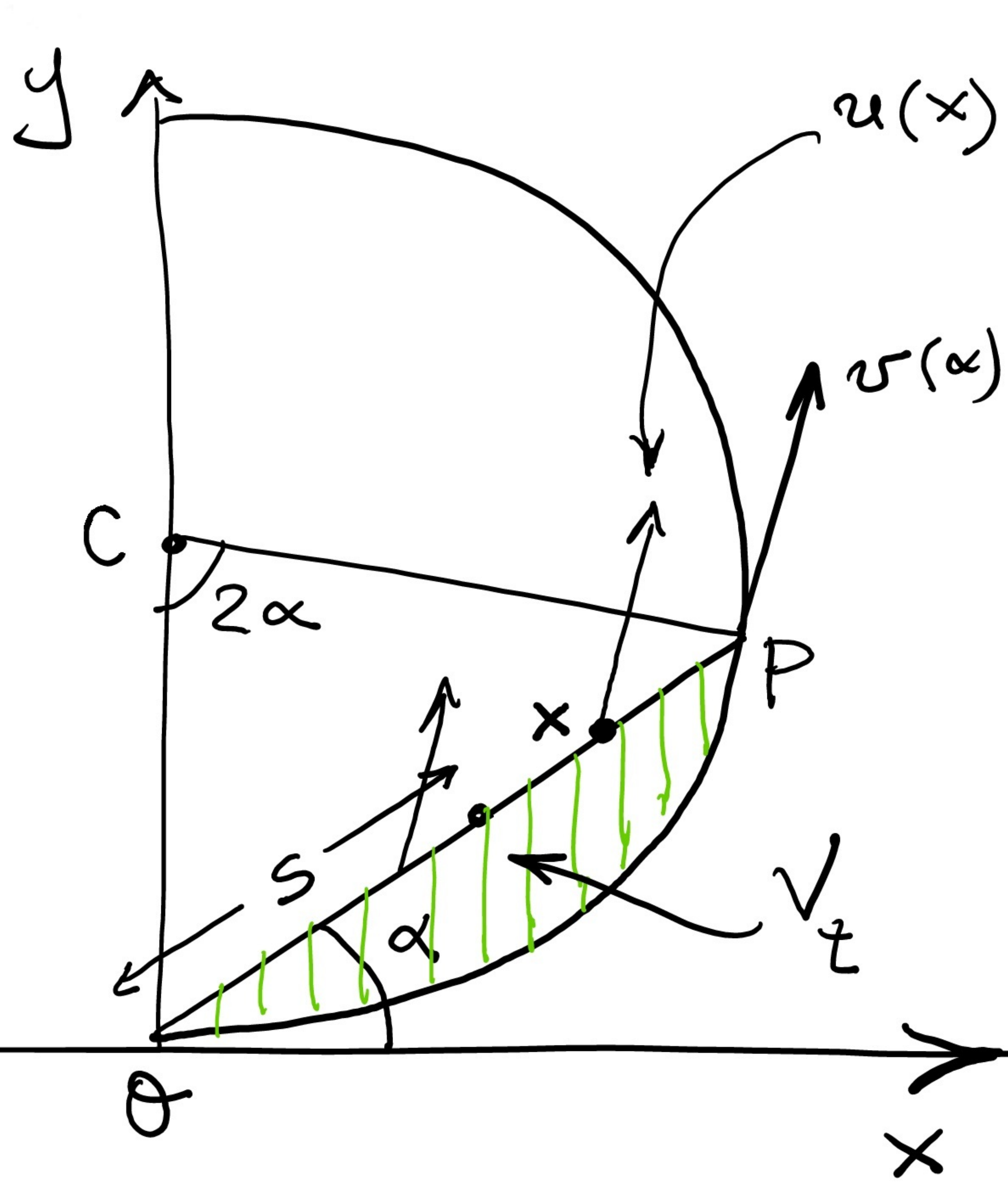},%
	{}]
	\noindent and the chord $\mathcal{O}P$, with $P$ moving along the circle with unit velocity $v(\alpha)$, $|v(\alpha)|=1$, and $\mathcal{O}$ fixed at the origin. The time evolution of $V_t$ can be put in the form $V_t=\phi_t^u (V_0)$ with $u(x)$ being the velocity of a point $x'$ on $\mathcal{O}P$ when it crosses $x$.
	By considering the angle $\mathcal O C P$ (where $C$ denotes the center of the circle) and some elementary geometry,
	it is easy to show that the length of $\mathcal{O}P$ is $2\sin \alpha$, so $u(s,\alpha)=s v(\alpha)/(2\sin \alpha)$, where polar coordinates $(s,\alpha)$ are being used. The 2-form $\omega$ we will consider is the area form, $\omega=s\td s\td \alpha$, with $\td \omega=0$, so that $I_t$ is just the area of $V_t$. For any (necessarily closed) 2-form $\eta=f(s,\alpha) \, \td s \td \alpha$ we have, by Cartan's
	formula,
\end{window}
\begin{align}
	L_u \eta
	=
	\td i_u \eta + i_{u} \td \eta
	=
	\td i_u \eta
	=
	\left(
	(u^1_{\phantom{1}1}+u^2_{\phantom{2}2})f+u^1 f_1+u^2 f_2
	\right) \td s \td \alpha
	\, ,
	\label{Luomega}
\end{align}
where subindices denote partials. Since $\omega$ and $u$ are both time-independent, (\ref{ptIt}) may be iterated to give
\begin{equation}
	\label{iterdiu}
	\partial_t^k I_t=\int_{V_t} (\td i_u)^k \omega
	\, .
\end{equation}
In cartesian coordinates $(x,y)$, $v(\alpha)=\cos(2\alpha) \partial_x+\sin(2\alpha) \partial_y$ (with $\alpha=\arctan\frac{y}{x}$).  Changing to polar coordinates $(s,\alpha)$, we get
\begin{equation}
	\label{usa}
	u(s,\alpha)=\frac{s}{2} \cot \alpha \, \partial_s+\frac{1}{2}\partial_\alpha
	\, .
\end{equation}
Putting $(\td i_u)^k \omega= s f_k \td s\td \alpha$, we find, after some algebra,
\begin{equation}
	\label{fk1234}
	f_0=1
	\, ,
	\quad
	f_1= \cot \alpha
	\, ,
	\quad
	f_2=\frac{1}{2}(\cot^2\alpha-1)
	\, ,
	\quad
	f_3=- \cot\alpha
	\, ,
\end{equation}
which implies all higher-$k$ results (since $f_3=-f_1)$.

Note that~(\ref{iterdiu}) can  be cast in the form (using Stokes' theorem)
\begin{equation}
	\label{iterdiu2}
	\partial_t^k I_t=\int_{\partial V_t} i_u (\td i_u)^{k-1} \omega
	\, .
\end{equation}
The integration is now along the boundary of $V_t$, consisting of the circular arc ${\mathcal{O}P}$, followed by the chord $\overline{P\mathcal{O}}$.
On the circular arc ${\mathcal{O}P}$, the integrand is zero, since the line element is along $u$, and the 2-form $(\td i_u)^{k-1}\omega$ has already been evaluated along $u$ in its first entry, so the only contribution to the integral is from the chord $\overline{P\mathcal{O}}$,
\begin{equation}
	\label{iterdiu4}
	\partial_t^k I_t=\int_{\overline{P\mathcal{O}}} i_u (\td i_u)^{k-1}
	\, .
\end{equation}
By (\ref{fk1234}) and (\ref{usa})
\begin{equation}
	i_u (\td i_u)^{k-1} = \frac{s}{2} f^{k-1} (s\cot \alpha \td \alpha- \td s)\, ,
\end{equation}
so the integral is,
\begin{equation}
	\label{iterdiu3}
	\partial_t^k I_t=\frac{s}{2}f^{k-1} \int_{s=0}^{2\sin\alpha} \td s \, s=
	\sin^{2} \alpha f^{k-1}
	\, .
\end{equation}
For the first three derivatives, (\ref{iterdiu3}) gives
\begin{equation}
	\label{area123}
	\partial_t I_t=\sin^2\alpha
	\, ,
	\quad
	\partial_t^2 I_t= \cos\alpha\sin\alpha
	\, ,
	\quad
	\partial_t^3 I_t= \frac{1}{2}-\sin^2\alpha
	\, .
\end{equation}
This is the correct result, as can be seen when comparing with the known result
$I_t=\alpha-\cos\alpha \sin\alpha$ that can be computed by geometrical means.
\bibliographystyle{ieeetr}
\bibliography{strings}
\end{document}